\documentclass[usenatbib,a4paper]{mn2e}
\usepackage[psamsfonts]{amssymb}
\usepackage{graphics}

\def\lap{$\buildrel < \over {_\sim}$} %less than or approx equal
\def\gap{$\buildrel > \over {_\sim}$} %greater than or approx equal
\def\trec{{t_{\rm rec}}}
\def\tmag{{t_{\rm mag}}}
\def\tob{{t_{\rm ob}}}
\def\rco{{R_{\rm co}}}

\def\sigcrit{{\Sig_{\rm crit}}} 
\def\sigcrith{{\Sig_{\rm crit}^{\rm high}}}

\def\mdisc{{M_{\rm ob}}}

\def\rmag{R_{\rm{mag}}}
\def\mdot{{\dot{M}}}

\def\ri{R_{{\rm i}}}
\def\riten{R_{{\rm i,10}}}

\def\rten{R_{{\rm 10}}}
\def\alphac{\alpha_{{\rm c}}}

\def\riten{R_{{\rm i,10}}}

\def\te{t_{{\rm e}}}
\def\rsun{{\rm R{{\rm_{\odot}}}}}

\def\msun{{\rm M_{\odot}}}

\def\eg{{e.g. \,}}

\def\etal{{et al.\,}}

\def\Sig{{ \Sigma}}
\def\pie{{ \pi}}
\def\nue{{ \nu}}
\def\Lambd{{ \Lambda}}

\def\mctt{\langle \mdot_{{\rm qu}} \rangle}
\def\mcttfif{\langle \mdot_{{\rm qu},15} \rangle}
\def\mfuo{\langle \mdot_{{\rm ob}} \rangle}

\def\lstar{L_{\star}}
\def\mstar{M_{\star}}
\def\lstardot{\dot{L}_{\star}}

\def\rstar{R_{\star}}
\def\tsp{t_{\rm{spin}}}
\def\tspo{t_{\rm{spin, ob}}}
\def\tspq{t_{\rm{spin, qu}}}
\def\trs{t_{\rm{rise}}}
\def\tth{t_{\rm{th}}}
\def\tvisc{t_{{\rm{visc}}}}
\def\te{t_{{\rm{e}}}}

\def\psp{P_{{\rm spin}}}
\def\vr{v_{\rm {R}}}

\def\bphi{B_{{\varphi}}}
\def\bz{B_{{z}}}
\def\omegak{\Omega_{\rm{k}}}
\def\omegastar{\Omega_{\rm{\star}}}
\def\rhod{\rho_{\rm{d}}}
\def\rc{r_{\rm{c}}}
\def\amag{a_{\rm{mag}}}

\begin{document}
\title[Outbursts of Young Stellar Objects]
{Outbursts of Young Stellar Objects}
\author[O.M. Matthews, R. Speith \& G.A. Wynn]{O.M. Matthews\footnotemark[1], R. Speith and G.A. Wynn\\
Department of Physics \& Astronomy, University of Leicester, 
Leicester, LE1 7RH \\}

\date{Received ** *** 2003; in original form 2003 *** **}

\label{firstpage}

\maketitle

\begin{abstract}
We argue that the outbursts of the FU Orionis stars occur on timescales
which are much longer than expected from the standard disc instability
model with $\alphac \gtrsim 10^{-3}$. The outburst, recurrence, and rise times are consistent with
the idea that the accretion disc in these objects is truncated at a
radius $\ri \sim 40 \  \rsun$. In agreement with a number of
previous authors we suggest  that the inner regions of the accretion discs in
FU~Ori objects are evacuated by the action of a magnetic propeller 
anchored on the central star. We develop an analytic solution for
the steady state structure of an accretion disc in the 
presence of a central magnetic torque, and present numerical
calculations to follow its time evolution. These calculations 
confirm that a recurrence time that is consistent with
observations can be obtained
by selecting appropriate values for viscosity and 
magnetic field strength.
\end{abstract}

\begin{keywords}
% Keywords selected from
% http://www.blackwell-science.com/~cgilib/jnlpage.asp?Journal=mnras&File=mnras&Page=authors/keywords (maximum of six)
accretion, accretion discs  -- stars: FU
Orionis -- stars: T Tauri -- stars: magnetic fields -- stars:
pre-main-sequence 
\end{keywords}

\footnotetext[1]{email address: owen.matthews@astro.le.ac.uk}

\section{Introduction}
\label{sec:int}
\label{sec:tta} 
The young stellar objects (YSOs) are stars which have not yet completed 
the process of star formation. They frequently retain some portion of 
their proto-stellar discs, from which mass accretion is thought to take place. 
The T Tauri stars are YSOs with masses $\lesssim 2$~${\rm M_{\odot}}$, 
and accretion rates in the region $\sim 10^{-7}-10^{-8}$~${\rm M_{\odot} yr}^{-1}$, 
\eg \cite{har97}. T Tau stars are divided into two subclasses, Classical T
Tau stars (CTTs) and Weak Line T Taus (WTTs). CTTs have been
shown to posses discs by the observation of double peaked absorption
lines, \eg \cite{har85}, and from the presence of an infra-red
excess \citep*{ada87,ken87}. CTTs have also been observed to produce extended emission \citep{bec84} which is consistent with the existence of an accretion disc.  WTTs are a faster spinning \citep{bou95} 
type of YSO with weak absorption lines and a less pronounced 
infra-red excess \citep{har98}. The presence of a disc is 
less certain in these systems.

Some YSOs have been observed to undergo outbursts, during which they
increase in brightness by up to two orders of magnitude on a 
timescale $\trs \sim$~1~yr. These
objects, termed FU Orionis stars, are otherwise very similar to CTTs.
Therefore it is generally assumed that FU Ori stars are members of the
same population as T Tau stars, \eg \citet{har85}. The outburst
duration has been estimated as $\tob$ \gap 50 yr \citep{har98},  
from the decay times of the outburst light curves (e-folding decay
times of the order of decades are typical). Both $\tob$
and the recurrence time of the outbursts ($\trec$) remain poorly
known as no stars have yet been observed to return to their
pre-outburst states. The peak outburst 
accretion rate is $\sim 10^{-4}$~${\rm M_\odot \: yr}^{-1}$, which is 
found by matching numerical simulations to outburst spectra, 
\eg \cite*{ken88}. The mean accretion
rate during outburst is $\sim 10^{-5}$~${\rm M_\odot \: yr}^{-1}$. 
For a detailed review of the FU Ori and T
Tau stars, and references, see \cite{har98,har96}.

The outbursts of YSOs are not yet fully understood,
but are believed to be associated with a thermal-viscous
accretion disc instability. Models based on the thermal-viscous disc
instability have widely been used as an explanation for similar
outbursts in other stellar systems, 
especially binary stars such as dwarf novae, \eg \citet{war95}. However 
the long recurrence times of YSOs do not correspond with those
predicted by the standard 
thermal-viscous disc instability model (DIM) with an accretion disc viscosity of the same order as that applied to binaries. In this paper we argue that the
standard DIM, with a quiescent Shakura-Sunayaev viscosity parameter of $\alphac \gtrsim 10^{-3}$, in the case of an accretion disc with an inner radius
comparable to the radius of the central star would predict a series
of shorter, more frequent outbursts than are observed in any YSOs. One possibility is that the observed
outburst behaviour could be reproduced simply by decreasing the Shakura-Sunyaev viscosity parameter to a very low value \eg \citet{bel94}. We propose an alternative model which 
incorporates the effect of a stellar magnetic field on the inner accretion disc.
A magnetic field anchored on a rapidly rotating star can act as
a propeller (\eg Lovelace, Romanova \& Bisnovatyi-Kogan 1999; Wynn, King \& Horne 1997) creating a low density region in the 
inner disc, and reducing accretion to the order of the CTT rate.
Outbursts are thought to begin when the surface density 
exceeds a critical value $\sigcrith$ anywhere in the disc. Since the 
critical density increases with distance from the star \citep*{can88}, 
a truncated accretion disc is able to accommodate more mass, 
forcing outbursts to be more dramatic and much less frequent.

In Section \ref{sec:met} we review estimates of the 
recurrence and rise times of outbursts in T Tau systems. These
estimates are compared with those expected from the standard DIM, we also outline the effect of viscosity. An
extension of the DIM, including a magnetic propeller,
is discussed in Section \ref{sec:new}. An expression for the evolution
of the disc in the radial dimension in the presence of a magnetic field is derived
in Section \ref{sec:one} and an analytic solution is obtained for
the steady state. Finally numerical calculations are 
performed which demonstrate the feasibility of using a magnetic
propeller to create the required depleted region in the centre of a T Tau disc.

%END OF SECTION

\section{Outburst Timescales}
\label{sec:met}

\subsection{Disc instability model}
\label{sec:dim}
The thermal-viscous disc instability model is now generally accepted as 
the mechanism behind outbursts in, for example, dwarf novae, \eg
(Buat-M\'{e}nard, Hameury \& Lasota, 2001; Warner, 1995), and other accreting binary systems.
However The standard DIM is unable to explain the long  
timescales associated with the outburst cycle of FU Ori stars at least for similar viscosities to those applied to discs in binaries. 

In the thermal-viscous instability model a limit-cycle
in temperature and density forces 
the accretion disc to cycle between two stable states: a hot, high
accretion, high viscosity state; and a cold, low accretion, low
viscosity state. Once an annulus of a disc reaches the hot state it
triggers a heating wave leading to an outburst which affects the
whole disc, or at least a portion of
the disc which extends to some maximum radius. The 
DIM is discussed in detail by \eg \citet*{fra01,
can88}. The trigger causing the change from the cold to the hot
state can be expressed in terms of a critical value of the disc surface 
density $\Sig$. Following \citet{can88} we approximate this critical $\Sig$ 
with the formula 
\begin{equation}\label{critic}
\label{crit}
\sigcrith = 11.4 \rten^{{\rm 1.05}} M_{1}^{{\rm -0.35}} {\it
\alphac}^{{\rm -0.86}} {\; \rm g \; cm^{-2}}\; ,
\end{equation}
where $\rten$ is the radius from the star in units of ${10^{10}}$
${\rm cm}$, $M_{1}$ is the mass of the star in solar masses, and ${\it
\alphac}$ is the Shakura-Sunyaev alpha viscosity during the cold
(quiescent) state. It is clear that the lowest
$\sigcrith$, which would trigger outbursts, must appear close to the inner
edge of the disc ($\ri$) and that outbursts would begin near  
$\ri$ unless some process were to cause a distortion of the disc's 
surface density profile. 
The radius at which the outburst is initiated has a strong effect 
on both $\tob$ and $\trec$.

It is possible to estimate the outburst duration $\tob$ 
by considering the e-folding decay time $\te$ of the observed light curves of FU Ori stars. 
These timescales are typically in the range of $10 {\rm{yr}} \lesssim \te \lesssim 50 {\rm{yr}}$.
 An FU Ori star in outburst is about 
two magnitudes brighter 
than in quiescence, \eg \cite{har98}. If the outburst is considered to be over when the
star returns to its quiescent luminosity then as a first approximation we
may say that the outburst time is of order $\tob \sim 0.4\te\Delta m \ln
10$ \lap $10^2$
yr, where $\Delta m$ is the change in magnitude between outburst and quiescence.

Since no FU Ori stars have been observed through a 
complete outburst cycle it is difficult to estimate $\trec$ from 
observational data. However it is possible to obtain a rough estimate 
of $\trec$ by comparing mean mass transfer rates in quiescence, 
$\mctt$, and during outburst, $\mfuo$. Assuming that a large proportion of the
region of the disc involved in the outburst is accreted, this mass 
must be replaced in a time $\trec$ and hence 
\begin{equation}
\label{compare}
\mfuo \tob \sim \mctt \trec 
\; .
\end{equation}
Solving relation (\ref{compare}) yields $\trec \sim 5000$ yr, 
where we have adopted $\mctt \sim 10^{-7}$~$\msun {\rm
yr^{-1}}$ and $\mfuo \sim 10^{-5}$~$\msun {\rm yr^{-1}}$ from
\citet{har98} and the estimate $\tob \sim 50$~yr. This estimate is
likely to be an upper limit as the mass transfer rate through the 
disc in quiescence must be larger than the observed accretion 
rate to allow an accumulation of mass leading to the next outburst.

A further approximate analytic limit for $\trec$ in the DIM can be obtained
by considering the mass flux through the annulus in the disc at which the outburst 
is initiated. The surface density of this region must reach $\sigcrith$
in a time $\trec$, leading to the relation
\begin{equation}
\label{arec1}
\trec \sim \frac{M(\ri)}{\mdot(\ri)} \sim \frac{2 \pi \ri \Sig \Delta R}{\mctt}
\; ,
\end{equation}
where $M(\ri)$ is the mass contained in the annulus, which begins 
close to the inner radius $\ri$ of the disc and extends outwards a distance $\Delta R$.  
Approximating $\Delta R \sim 0.1 \ R$ (which is not unreasonable if
an aspect ratio of $H / R \sim 0.1$ is adopted throughout the disc)  
it is possible to express $\trec$ in the form,
\begin{equation}
\label{arec2}
\trec \sim \frac{0.2 \pi \ri^2 \Sig}{\mctt}
\; .   
\end{equation}
During quiescence the surface density in the disc is limited by $\Sig < \sigcrith$. 
Therefore equations (\ref{arec2}) and (\ref{crit}) can be combined to give
\begin{equation}
\label{arec3}
\trec \lesssim  2 \times 10^{-2} M_{1}^{-0.35} \riten^{3.05} \mcttfif^{-1} \alphac^{-0.86} \  \rm{yr} 
\; ,
\end{equation}
where $\riten$ represents the inner radius of the disc in units of $10^{10}$~cm, 
$\mcttfif$ is the mean quiescent accretion rate in units of $10^{15} \  \rm{gs^{-1}}$. 
In this paper we adopt a value of $\alphac \sim 0.01$ which is the quiescent viscosity parameter 
commonly applied to dwarf novae accretion discs. This assumption will be discussed further in section \ref{sec:visc}. 
The estimate of the surface density profile that is used here is of course an upper limit, it will be seen however in Section \ref{sec:one} that this is a reasonable approximation. 
The estimate (\ref{arec3}) can be tested by comparing its predictions to the
outburst behaviour of the dwarf novae, which have been observed
through many outburst cycles.  In a dwarf nova the accreting
star is a white dwarf with a mass $\mstar \sim 1 \  \msun$, and radius $\rstar \sim 0.05 \
\rsun$, and the quiescent accretion rate is $\mctt \lesssim 10^{-11} \ \msun {\rm{yr^{-1}}}$ \citep{war95}. Using these values, and assuming that $\ri \sim \rstar$, 
equation (\ref{arec3}) gives an estimate for the recurrence time of 
$\trec \sim 10$ d, very close to the observed $\trec$ 
in these systems.
In order to apply (\ref{arec3}) to a typical FU Ori system we  
adopt values of $\mstar \sim 1 \  \msun$, and $\mctt \sim 10^{-7} \  \msun {\rm{yr^{-1}}}$. 
If the disc is 
assumed to extend down to a typical CTT stellar radius of $\ri \sim 2 \rsun$, then equation 
(\ref{arec3}) yields a recurrence time of $\trec \sim 0.5 \  \rm{yr}$ 
This is clearly far too short to be consistent
with observation. Indeed, to obtain $\trec \sim 5000 \  {\rm{yr}}$ we
require an inner disc radius of $\ri \sim 40 \  \rsun$.  

The timescale on which the luminosity of the disc increases ($\trs$)
in an FU Ori outburst is governed by the thermal timescale of the 
region in which the outburst begins, in the case of an outside-in outburst.
 \citet{cla96} have examined this is detail and point out that rise times of \lap 1 yr imply that outbursts
must begin at radii much larger than that of the central star. They estimate values \gap 15 $\rsun$, in rough agreement with the
estimates above. If we assume the observed values of $\trs$
are of the same order as the local thermal timescale $\tth$ at the point 
where the outburst begins $\tth$ we obtain the relation \citep{fra01}
\begin{equation}
\label{thermone}
\trs \sim \tth \left( R \right) \sim \left( \alpha \Omega_k \left( R \right)
\right)^{-1} \sim \frac{1}{\alphac}\left( \frac{R^3}{G \mstar} \right)^{1/2}
\; ,
\end{equation}
where $\Omega_k \left( R \right)$ represents the Keplerian angular
velocity at a given radius. The alpha viscosity parameter in
quiescence is once again assumed to be $\alpha \sim 0.01$. 
The observed rise time of $\trs \sim 1 \  {\rm{yr}}$ is consistent with an inner disc radius of $\ri \sim 40 \  \rsun$, which is similar to the estimate obtained from $\trec$ and consistent with the large radii predicted by \cite{cla96}.

\subsection{Low viscosity models}
\label{sec:visc}
It has been shown that the above estimates for $\trec$ and $\trs$ are 
consistent with the DIM if it is modified, for example, by including 
a truncated inner disc where $\ri \sim40 \  \rsun$. We note here however 
that long values of $\trec$ and $\tob$, consistent with
observation, can be reconciled with the standard DIM if a low 
value of the quiescent disc viscosity is invoked. Equation (\ref{arec3})
has a dependence on $\alpha_c$ and so reducing $\alphac$ to small values 
can yield a $\trec$ which is arbitrarily long. Indeed such a model has 
been applied by \citet{bel94} and \citet{bel95} who use a one dimensional disc model similar to that described in Section \ref{sec:one}. \citet{bel94} simulate full outbursts and include detailed thermodynamics in order to calculate the onset of the thermal instability. This is then associated with an arbitrary increase in the disc viscosity. For most of these simulations $\Sigma = 0$ at the inner boundary and $\mdot$ is held constant at the outer boundary. With this model \cite{bel94} are able to obtain recurrence times of $\trec \sim 1000 \ {\rm yr}$ using the standard DIM with $\alphac \sim  10^{-4}$. Our analytic estimate of $\trec$ from Equation (\ref{arec3}) requires an even lower value of $\alphac \sim 10^{-6}$ to obtain $\trec$ of this order. This discrepancy is because the analytic estimate developed in \ref{sec:dim} includes the assumption that the outburst begins at a radius of $R = R_{\rm{i}} \sim 2 \rsun$ whereas according to the results of \cite{bel94} the outburst begins at $R \sim 10 \rsun$. \citet{bel94} also include a feedback mechanism whereby the radius of the disc is a function of $\mdot$. We note here that this model is successful, providing that outbursts can be initiated at $R \gtrsim 10 \ \rsun$. Of course a similarly long $\trec$ could be replicated by outbursts initiated at smaller radii assuming even lower viscosities. Nonetheless \cite{bel94} do confirm that the dwarf nova like estimate of $\alphac \sim 10^{-2}$ is far too high to reproduce values of $\trec$ in agreement with observational limits in the standard DIM. 

Theoretical work has been done which suggests that a very low value of $\alphac$ is possible in protostellar discs, because the region in which thermal ionisation is able to occur throughout the thickness of the disc is limited to $R \lesssim 0.1 \ {\rm{AU}}$ \citep{gam96}. However the FU Orionis outbursts are driven in the inner disc, which is likely to be thermally ionised, and where as a result the viscosity is likely to be similar to that in dwarf novae. \citet{har97} have argued that the radii of circumstellar discs can be used to constrain the quiescent viscosity parameter to $\alphac \sim 0.01$, similar value to the dwarf nova case, throughout the disc. The remainder of this paper will be dedicated to the investigation of an alternative mechanism which can extend recurrence times, by disc truncation, to within the observational limits while using a dwarf nova like low state viscosity of $\alphac \sim 0.01$ in the inner disc.

%END OF SECTION

\section{Disc truncation by a magnetic field}
\label{sec:new}

In Section 2 we argued that the standard thermal-viscous instability
outburst model is insufficient to explain the outbursts of the FU Ori
stars. A truncated disc however produces reasonable estimates of
$\trec$ and $\trs$. One possible mechanism for truncating the disc is
a magnetic field associated with the accreting star. 

The suggestion that the accretion disc in YSOs is truncated by the
magnetic field of the young star, or otherwise, is not new and has
appeared in the literature many times. Examples include \citet*{ult01}, who
model the CTT infra-red excess with a truncated disc, \citet{arm96}
who explain T Tau spin regulation with a truncated disc model, and \citet{cla96}.  A magnetic mechanism which may 
lead to disc truncation is investigated below.

\subsection{The magnetic propeller}
\label{sec:mag}
The magnetic fields of the T Tau and FU Ori stars are widely believed
to play an important role in their evolution and behaviour, \eg
\citet{yi97,saf99}. 
Good estimates of the magnitudes of these fields
are, however, difficult to obtain. The most recent estimates available
of the magnetic field of the CTT star BP Tau, measured using
spectropolarimetry \citep{joh99}, put the value at $B = 2.4 \pm
0.1$~kG. This is in good agreement with earlier measurements for the
magnetic fields of both CTTs and FU Ori stars, which have been
calculated using Zeeman broadening effects, \eg \cite{joh87}. 
The inner accretion disc in a CTT will be partially ionised,
as it is irradiated by the accreting star, and will therefore interact
with the stellar magnetic field. Angular momentum can be transfered from the
disc to the star and vice-versa, via this magnetic interaction. The
magnetic field is therefore important in determining the spin rates
of these stars as discussed in more detail in \cite{pop96,arm96}. 

The magnetic propeller effect (\eg Wynn \etal 1997; Lovelace \etal 1999) involves a net 
transfer of angular momentum from the star to the disc via
the magnetic field. In this case the interaction between the 
disc and the field takes place at radii larger than the co-rotation
radius $\rco$, the point at which the circular Keplerian angular
frequency $\Omega_K$ of the disc matches the angular frequency ${\it \Omega_\star}$ of the star. 
For the purposes of this paper we assume that the
stellar field is dipolar and that it rotates at the same rate as 
the star such that
\begin{equation}\label{angmom}
{\it \Omega}_K(\rco) = {\it \Omega}_{\star} = {\it \Omega}_{B}
\; ,
\end{equation}
where ${\it \Omega}_{B}$ represents the angular frequency of the
magnetic field. At
$\rco$ it is clear that a body in a circular Keplerian orbit remains
stationary in the frame of reference of the magnetic field, and there 
is no torque between the disc and the field at this point.
Hence, the magnetic timescale $\tmag$ (the timescale
on which magnetic torques redistribute angular momentum in the disc) 
is infinite at $\rco$. Outside $\rco$ we have $\Omega_K < \Omega_\star$
and the disc gains angular momentum at the expense of the star. 
Matter which is accelerated in
this manner will be pushed out to greater radii until the magnetic
torque (which for a dipole field decreases as $\sim R^{-6}$) becomes 
small when compared to viscous stresses in
the disc. This occurs when $t_{\rm{mag}}\gg t_{\rm{visc}}$, where
$t_{\rm{visc}}$ is the timescale on which viscous stresses
redistribute angular momentum within the disc. The radius at which 
$\tmag \sim \tvisc$ is usually termed  the magnetic or magnetospheric radius $\rmag$.
Inside $\rco$, the magnetic torque extracts angular momentum from the
disc. In this case the disc mass is rapidly accreted onto the star. 

The results of both the propeller and the accretion regimes are the
formation of a depleted region in the disc close to the star. The
radius of this region is $\ri \sim \rmag$, and therefore is 
a function of $B$, the spin period $\psp$ of the star, and the disc viscosity. If we assume, as
argued in Section \ref{sec:met}, that an inner disc radius 
$\ri \sim 40 \ \rsun$ is required to 
be consistent with observational constraints then $\ri$ must 
exceed $\rco$ even for slow spinning T Taus. Although both the 
accretion and propeller regimes lead to the formation of a depleted
region in the centre of the disc, only the propeller mechanism leads to the 
accumulation of disc mass at radii greater than $\rco$.

Below we derive an expression for the magnetic timescale $\tmag$
in a completely magnetised accretion disc in the presence of a
dipolar magnetic field. The result is almost identical to that adopted
by \citet{liv92}. The inclusion of
a magnetic field in a hydrodynamic system introduces two additional
terms to the Euler equation; a magnetic pressure term and a magnetic 
tension term, \eg \cite{den90}. The magnetic pressure term is
negligible where $B$ is small. The magnetic tension term can be
expressed as 
\begin{equation}\label{btscl1}
\amag \sim \frac{1}{\rhod \rc} \left( \frac{\bz \bphi}{4 \pie} \right) \hat{\mbox{\boldmath$v$}}
\; ,
\end{equation} 
where $\rhod$ represents the density of the disc, $\rc$ is the local
radius of curvature of the field lines and $\bz$ and $\bphi$ represent
the vertical and azimuthal components of the magnetic field
respectively. We use the approximations $\rc \sim H \bz / \bphi$
\citep*{pea97}, and  
$H \sim 0.1 R$. The ratio of field strengths can be expressed in the form (\eg Livio \& Pringle 1992)
\begin{equation}\label{btscl2}
\frac{\bphi}{\bz} \sim - \frac{\left(\omegak - \omegastar \right)}{\omegak}
\; .
\end{equation}
The magnetic timescale can be defined in terms of the magnetic
acceleration and the Kepler velocity by the relation 
\begin{equation}
\tmag \sim \frac{R \omegak}{\amag} \sim - \frac{2 \pie R^{2} \rhod}{5
  \bz^{2}} \frac{\omegak^{2}}{\left({\omegak-\omegastar}\right)}
\; . 
\end{equation}
Finally the volume density can be related to the surface density by
the expression $\rhod \sim \Sig / H \sim 10 \Sig / R$ and, for a
dipole field, we have $\bz \sim \mu R^{-3}$ where $\mu$ is the magnetic moment of the
star. These relations lead to the following expression for the magnetic timescale 
\begin{equation}
\label{tmagspec}
\tmag \sim \frac{4 \pie \Sig \sqrt{G \mstar} R^{11/2}}{\mu^{2} \left( \left[\frac{R}{\rco} \right]^{3/2} - 1 \right)}
\; .
\end{equation}   
It is of course possible to use other prescriptions for the magnetic
timescale, for example the diamagnetic case as used by \eg
\citet{wyn97}. In general however $\tmag$ can normally be parametrized
in the form
\begin{equation}
\label{tmaggen}
\tmag \sim \frac{2 \Sigma}{\beta} \frac{R^{\left( \gamma+2 \right)}}{\left( \left[\frac{R}{\rco} \right]^{3/2} - 1 \right)}
\; ,
\end{equation}
where $\gamma$ and $\beta$ are parameters determined by the magnetic
interaction model. We use this general expression to derive equation (\ref{genadvec}) in
Section~\ref{sec:one}. In the fully magnetised case represented by
equation (\ref{tmagspec}) we have $\gamma = 7/2$ and $\beta$ is
determined by the mass and magnetic moment of the star
according to 
\begin{equation}
\label{betalast}
\beta = \frac{{\mu^{2}}}{2 \pi \sqrt{G \mstar}} 
\; .
\end{equation}

\subsection{Magnetic models for CTTs and WTTs}
\label{sec:ctmod}

Several authors have considered magnetic truncation of the accretion
disc in YSOs. \citet*{ken95} use this idea to model the infrared
excesses of CTTs. \citet{arm96} model a magnetically truncated
accretion disc and use this model to explain the regulation of the
spin periods of T Tau stars on a timescale $\sim 10^{5} \  \rm{yr}$. 
The magnetic propeller effect has previously been applied to
the case of T Tau accretion discs by \citet{ult01}, who use a
smoothed particle hydrodynamics technique to show that a central
hole is formed by the magnetic field, exactly as described above and
as will be confirmed by numerical results in Section \ref{sec:one}. In
the full three dimensional case some accretion should occur, as there
will be a component of the orbital velocity which is not in the
azimuthal direction with respect to the dipole field. The accretion
stream follows the magnetic field lines of the accreting star. Indeed
\citet{gul98} find that CTT spectra agree with an accretion stream of
this sort better than they do with standard boundary layer
accretion. This is analogous, in binary systems, to the flow in the
case of the intermediate polar (IP) stars as discussed for example in
~\cite{mur99}. \citet{liv92} discuss the effect of a magnetic
propeller on the onset of outbursts in dwarf novae.

Magnetic models have also been used to explain the transition between CTT
and WTT states by allowing the magnetospheric radius $\rmag$ to move,
which affects the accretion behaviour of the system. When $\rmag$ is
large, the disc is propelled outwards to large radii and very little
accretion takes place. This state can be identified with a WTT system
since there would be a low accretion rate and no inner disc. If the
star is in a state where $\rmag$ is smaller, then it may be that
$\rmag < \rco$, in which case the system would become a magnetic
accretor. As an alternative to the magnetic accretor, a weak propeller
could be invoked. In this case $\rmag > \rco$ but $\rmag$ is still
smaller than the strong field value. In either case these systems
could then be identified as CTTs as a disc would exist close to the
star, although not extending to the surface. This disc could produce
the observed infra-red excess and emission line characteristics of
CTTs. 

\citet{cla95} suggest a `magnetic gate' where magnetic cycles in the
star cause a change in the magnetospheric radius, $\rmag$. A second
possibility is that proposed by \citet{arm96} where $\rmag$ can be
changed by spinning the star up or down by disc interactions. High
spin stars, with a greater $\rmag$, would then be classified as WTTs,
which is in agreement with observation. A rapid spin up can most
simply be achieved by a high accretion phase such as an FU Ori
outburst. The gradual spin down of the system in its WTT phase and
transition to CTT is modelled in detail by \citet{arm96}. The effects
of a magnetically truncated disc on the spin evolution of the T Tau
stars will be considered in Section \ref{sec:spev}.

\subsection{A magnetic model for FU Orionis outbursts}
\label{sec:mod}
As discussed in Section \ref{sec:ctmod}, the action of a magnetic
propeller in the disc of a T Tau object will cause the truncation of
the disc at some point outside $\rco$. The precise position of truncation
depends on the strength of the stellar magnetic field, the rate of
spin of the star and the viscosity of the disc. If a depleted region, or
hole, is formed then the total mass $\mdisc$ of the disc is allowed to
reach a higher value before $\sigcrith \left( \ri \right)$ is exceeded
and an outburst begins, increasing both $\tob$ and $\trec$. During the quiescent, or propelling, phase mass transfer is not completely prevented, but is reduced to the CTT accretion rate of $\mdot \sim 10^{-7} \ {\rm \msun yr^{-1}}$. Some accretion will still occur because in three dimensions, as mentioned in Section \ref{sec:ctmod}, some component of the azimuthal velocity is not perpendicular to the magnetic field. The mass which leaks through the magnetic barrier accretes along the magnetic field lines onto the central star, which is analogous to the behaviour of other magnetic propeller systems. In a forthcoming paper we attempt to apply a similar magnetic mechanism to the cataclysmic variable WZ Sagittae, in which case the quiescent accretion rate is $\sim 1$ percent of the outburst rate. Indeed even in the interacting binary AE Aquarii \citep{wyn97}, a propeller system so strong that it completely prevents the formation of a disc, the accretion rate is $\sim 1$ percent of secondary mass loss rate. It should be noted that this accretion is an intrinsically three dimensional effect and that the situation is somewhat different in the one dimensional model applied later in the paper. This distinction is discussed in section \ref{sec:cod}. 

It is of course possible to imagine a disc in which the propeller is sufficiently strong to prevent outbursts. In this case a steady state may exist in which the accretion rate is constant and the critical density is never reached. The parameters for which this is possible will be investigated in section \ref{sec:res}. It should be stressed here however that in order for any steady state to exist, some `leakage' of material through the magnetic barrier is required. The relative frequency of outbursting and non-outbursting CTTs will depend on the distribution of magnetic moments and of disc viscosities. The discussion of this effect requires a full parameter search and is left to future work. 
 
If the disc goes into outburst, the viscosity 
is believed to increase by up to an
order of magnitude. The resulting decrease in $t_{\rm{visc}}$ 
reduces $\rmag$ to a point much closer
to, or even within, the surface of the star, 
and causes $\mdot$ to increase by
several orders of magnitude to the peak FU Ori accretion rate of $\mdot \sim 10^{-4} \ {\rm \msun yr^{-1}}$. The dissipative (viscous) effect dominates over the
advective (magnetic) effect at all radii in this case. This continues
to be the case until much of the stored mass in the disc has been
accreted and the disc is allowed to return to the cold state. Indeed
for as long as the condition $\ri < \rco$ is satisfied, the magnetic
field will actually assist accretion in the inner disc.  

On the disc returning to the cold state, $\ri$ is forced out once more
to its original position outside $\rco$, assuming that neither $B$ nor $\psp$ are
significantly altered by the outburst. The system should now return to
its pre-outburst state, with a CTT accretion rate, and the mass of the outbursting region should
begin to build up once more. Hence, providing that a magnetic propeller can
produce a depleted region, it can be invoked to
explain the long $\trec$ as well as the long $\trs$ in FU Ori systems.

\subsection{Spin evolution}
\label{sec:spev}
Since $\psp$ effects $\rmag$ and hence has an
influence on $\trec$, it is important to consider the effect
of spin evolution on FU~Ori outbursts. This can be done to a first
approximation by comparing the spin up timescale of the star to
$\trec$ and $\tob$.

According to \citet{arm96} the squared radius of gyration of a T Tau
star is $k^{2} \sim 0.2$ for most of its lifetime. We adopt this value
of $k^{2}$ and assume a solar mass star with a typical T Tau spin
period of $\psp \sim 3$~d. The rotational angular momentum of the star
is given by 
\begin{equation}
\label{spintwo}
\lstar = k^{2} \mstar \rstar^{2} \Omega_{\star}
\; .
\end{equation} 
Away from the inner radius angular momentum transport in the accretion 
disc is dominated by viscous effects. It is only
at the inner edge of the disc that magnetic effects
become important. 

During quiescence the magnetic propeller will cause the star to 
experience a net spin down torque. \citet{arm96} suggest that
a propeller will spin down the star on a timescale of
$\tspq \sim 10^{5} \ \rm{yr}$. It is likely then that 
stellar spin evolution is insignificant during a single quiescent phase.

In the case of magnetic accretion the spin up
torque on the star is given by
$\lstardot = \mdot l \left( \ri \right)$, where $l \left( \ri
\right)$ represents the specific angular momentum of matter at
$\ri$, giving 
\begin{equation}
\label{spinone}
\lstardot = \mdot \sqrt{G \mstar} \ri^{1/2}
\; .
\end{equation}
The spin up timescale can be expressed as 
\begin{equation}
\label{tspin}
\tsp \sim \lstar / \lstardot 
\; .
\end{equation}
During outburst $\mfuo \sim 1 \times 10^{-5} \ {\rm M_\odot
yr}^{-1}$ and $\ri \sim \rstar \sim 2 \   \rsun$ because the central
region is likely to have become filled. From equation (\ref{tspin}), 
this yields
$\tspo \sim 10^{3} \ \rm{yr}$.  Therefore even in outburst
$\tspo > \tob$ compared with an estimated $\tob \sim 10^{2} \
\rm{yr}$. In this case it is sensible to assume that the measured $\mdot$ is
the same as that at the surface of the star and to neglect transport of 
angular momentum from the star to the disc. 
This result shows that spin evolution should therefore not have a 
significant influence on outburst behaviour during one outburst, although 
there may be a measurable cumulative effect.

Although spin evolution may turn out to be significant in the longer
term for T Tau stars, it is still possible to illustrate the principle
of outburst and recurrence time behaviour with a fixed spin and hence
a fixed $\rmag$. It is assumed for the purposes of this paper that the
outbursts occur in the CTT phase and that the system returns directly to a
similar CTT state after outburst. Any magnetic evolution of the T Tau star 
is neglected.

%END OF SECTION

\section{The Structure of a truncated disc}
\label{sec:one}

\subsection{Analytic Solution}
\label{sec:cod}
A thin, axisymmetric accretion disc is assumed to be rotating about a
central mass $\mstar$. As we are only concerned with the inner
Sections of the disc where outbursts begin, and the total disc mass
can be estimated to be $\sim$10$^{-3}$~$\msun$ \citep{har98}, it is
reasonable to neglect self-gravity in this portion of the
disc. Averaging the basic hydrodynamic equations over all azimuthal
phases and integrating over the direction normal to the plane of the
disc, the continuity equation
becomes
\begin{equation}\label{conteq}
\frac{\partial\Sig}{\partial t} + 
\frac{1}{R}\frac{\partial(R\Sig v_{\rm R})}{\partial R} = 0
\; ,
\end{equation}
the radial component of the Navier Stokes equation becomes
\begin{eqnarray}\label{radnavsto}
\lefteqn{\Sig\left(
\frac{\partial v_{\rm R}}{\partial t} + v_{\rm R}\frac{\partial v_{\rm R}}{\partial R}
- \frac{v_\varphi^2}{R}
\right)  = 
{}-\frac{\partial p}{\partial R} 
- \Sig\frac{GM}{R^2}}
\\
\nonumber
 & &
{}+
\frac{4}{3R^{3/2}}\frac{\partial}{\partial R}
\left[R^{3/2}\nue\Sig\frac{\partial v_{\rm R}}{\partial R}\right]
-
\frac{2}{3R^3}\frac{\partial(R^2\nue\Sig v_{\rm R})}{\partial R}
\; ,
\nonumber
\end{eqnarray}
and the azimuthal component of the Navier Stokes equation becomes
\begin{equation}\label{azinavsto}
\Sig\left(
\frac{\partial v_\varphi}{\partial t} + 
\frac{v_{\rm R}}{R}\frac{\partial(R v_\varphi)}{\partial R}
\right) =
\frac{1}{R^2}\frac{\partial}{\partial R}\left[
R^3\nue\Sig
\frac{\partial}{\partial R}\left(\frac{v_\varphi}{R}\right)\right]
\; ,
\end{equation}
where $v_{\rm \varphi}$ and $v_{\rm R}$ denote the azimuthal and radial
components of the velocity respectively and $\nue$ represents the
kinematic viscosity of the disc. 

By multiplying the azimuthal component (\ref{azinavsto}) of the
equation of motion with $R$, we obtain an equation for the specific
angular momentum in the disc,
\begin{equation}\label{angnavsto}
\Sig\left(
\frac{\partial l}{\partial t} + 
v_{\rm R}\frac{\partial l}{\partial R}
\right) =
\frac{1}{R}\frac{\partial}{\partial R}\left[
R^3\nue\Sig
\frac{\partial}{\partial R}\left(\frac{l}{R^2}\right)\right]
+ \Sig\Lambd
\; .
\end{equation}
Here, we have already assumed an additional external torque acting on
the disc, where $\Lambd$ is the injection rate of angular momentum per
unit mass, following the same procedure as \cite{lin86}. A similar
procedure is also employed by \cite{pri91} in the case of a binary
system.

Assuming a sufficiently cold disc and therefore that the dynamical
timescale $t_\mathrm{dyn}\sim R/v_\varphi$ is much smaller than any
other timescale, we can adopt a Keplerian approximation for the
azimuthal motion such that $v_\varphi = (GM/R)^{1/2}$. Then,
(\ref{angnavsto}) can be solved for the radial velocity in the disc,
\begin{equation}\label{radvel}
v_{\rm R} =
-\frac{3}{R^{1/2}\Sig}
\frac{\partial(R^{1/2}\nue\Sig)}{\partial R}
+ 2\Lambd\frac{R^{1/2}}{\sqrt{GM}}
\; .
\end{equation}
Inserting equation (\ref{radvel}) into the continuity equation
(\ref{conteq}) results in an evolution equation for the surface
density of the disc,
\begin{equation}\label{sigmaevol}
\frac{\partial\Sig}{\partial t} =
\frac{3}{R}\frac{\partial}{\partial R}\left[
R^{1/2}
\frac{\partial(R^{1/2}\nue\Sig)}{\partial R}
\right]
- 
\frac{1}{R}\frac{\partial}{\partial R}\left[
2\Lambd\Sig\frac{R^{3/2}}{\sqrt{GM}}
\right]
\; ,
\end{equation}
where the right hand side is composed of a diffusion term on the left
and an advection, or torque term on the right.

To parametrize the specific torque $\Lambd$, we write
\begin{equation}\label{torque}
\Lambd = \frac{l}{t_{\Lambd}} = \frac{\sqrt{GM}R^{1/2}}{t_\Lambd}
\; ,
\end{equation}
where $t_\Lambd$ is the timescale on which the local disc material
gains angular momentum.

In the present case, the source of the torque is the magnetic
interaction of a rapidly rotating, moderately magnetic YSO within the
disc. The torque timescale is therefore equivalent to the magnetic
timescale, $t_\Lambd\sim t_\mathrm{mag}$. This magnetic timescale is
given in equation (\ref{tmaggen}).

In general, the kinematic viscosity $\nue$ may depend on radius as
well as on surface density, $\nue = \nue(R,\Sig)$. For the purposes of
this paper $\nue$ is taken to be independent of $\Sigma$ and
parametrized according to a power law such that $\nu = \delta
R^{\epsilon}$. Kinematic viscosity may also change with time according
to an outburst limit cycle. It may be possible to implement this form
of $\nue$ and, with suitable boundary conditions, to simulate complete
outburst cycles in the future. The viscous timescale
is given by \citep{fra01}
\begin{equation}
\label{tscale2}
t_\mathrm{visc}\sim 
\frac{R^{\rm 2}}{\nue}\sim
\frac{R^{\rm 2}}{\delta R^{\epsilon}}
\; .
\end{equation}
By considering equation (\ref{tmaggen}), a general expression for the
advection term in equation (\ref{sigmaevol}) is obtained
\begin{equation}\label{genadvec}
\frac{1}{R}\frac{\partial}{\partial R}\left[
2\Lambd\Sig\frac{R^{3/2}}{\sqrt{GM}}
\right]
=
\frac{{\it \beta}}{R}\frac{\partial}{\partial R}\left[
\frac{1}{R^{\gamma}}
\left(\left[\frac{R}{R_\mathrm{co}}\right]^{3/2} - 1\right)
\right]
\; .
\end{equation}
Equation (\ref{sigmaevol}) can be solved analytically for the steady state.
By definition in the steady state $\mdot = - 2 \pi R \Sigma \vr$ is a constant
throughout the disc. Using equation (\ref{radvel}) and with a magnetic torque 
as defined in equation (\ref{torque}) this condition can be used to derive a 
general solution for $\Sigma$ in steady state,
\begin{eqnarray}\label{gensignew}
\lefteqn{\Sigma = {} \frac{\mdot}{3 \pi \delta} R^{-\epsilon}+
  \frac{\beta R^{-\left(\gamma + \epsilon \right)}}{3 \left(2 - \gamma
    \right) \delta}
\left(\left[\frac{R}{\rco} \right]^{3/2} - \frac{\left(2-\gamma \right)}{\left(\frac{1}{2} - \gamma \right)} \right)}
\\
\nonumber
 & &
{}+ 
\frac{C}{\delta} R^{- \left(\frac{1}{2} + \epsilon \right)}
\; ,
\nonumber
\end{eqnarray}
where C is an arbitrary constant. As will be seen this equation has a form such that 
$\Sigma$ becomes negative close to the star. The inner part of this steady
state solution is thus clearly unphysical. It is difficult therefore for a physically 
motivated inner boundary condition to be chosen. Furthermore the general shape of the 
function is rather sensitive to this boundary condition. 
The outflow condition $\partial \Sigma / \partial R = 0$ 
will be applied however as a first approximation. The inner boundary is taken to be very close to
the surface of the star. This gives, in its most general form,
\begin{eqnarray}\label{cbound}
\lefteqn{C
= {}
\frac{\beta \rstar^{-\gamma}}{3 \left(\frac{1}{2} + \epsilon \right)} \left( \frac{\left( \gamma + \epsilon  \right)}{ \left( \frac{1}{2} - \gamma \right)} 
\rstar^{1/2}  
+\frac{\left( \frac{3}{2} - \gamma - \epsilon \right)}{\left(2 - \gamma \right)} 
\frac{\rstar^{2}}{\rco^{3/2}}\right)}
\\
\nonumber
 & &
{}-
\frac{\epsilon \mdot \rstar^{1/2}}{3 \pie \left( \frac{1}{2} + \epsilon \right)}
\; .
\nonumber
\end{eqnarray}
This analytic solution for the steady state is precisely reproduced by the numerical solution in 
the following Section. It is however instructive to examine the analytic form of the solution. 
It can be seen from the form of equation (\ref{gensignew}) that, counter-intuitively, in steady state the 
position of $\ri$,  defined here as the point where $\Sigma = 0$, is independent of $\delta$. 
The disc structure at very large radii tends to the non-magnetic case as would be expected. 
The non-magnetic solution is dependent only on $\mdot$,  and $\nu$.
It is almost identical to the solution for the steady thin disc presented by \citet{fra01}. Indeed 
this solution, 
\begin{equation}
\label{fkr}
\nu\Sigma = \frac{\mdot}{3 \pi} \left( 1 - \left[ {\frac{\rstar}{R}} \right]^{1/2} \right) 
\; ,
\end{equation}
can be reproduced precisely by applying an inner boundary condition such that $\Sigma = 0$.
In this case, as well as in the full magnetic case it holds that $\Sigma \propto 1/\nu$.

The steady state solution in one dimension cannot represent a physical situation. In steady
state $\mdot$ is constant throughout the disc by definition. However in the steady state 
solution found above there is an evacuated region in the disc through which no matter can pass.
Numerically this represents the transport of negative mass, but this clearly has no physical
meaning. In three dimensions however it is possible to imagine similar behaviour by allowing 
mass to flow along magnetic field lines in the inner disc. The one dimensional solution can then be
taken as an approximation of the behaviour of matter in the plane of the disc. This approach
is given some validity by the similarity of these one dimensional results to those obtained in
three dimensional simulations. This can be seen in \eg \citet{ult01} and also in our own preliminary three dimensional computations.

%END OF SECTION

%\begin{eqnarray}\label{radnavsto}
%\lefteqn{\Sig\left(
%\frac{\partial v_{\rm R}}{\partial t} + v_{\rm R}\frac{\partial v_{\rm R}}{\partial R}
%- \frac{v_\varphi^2}{R}
%\right)  = 
%{}-\frac{\partial p}{\partial R} 
%- \Sig\frac{GM}{R^2}}
%\\
%\nonumber
% & &
%{}+
%\frac{4}{3R^{3/2}}\frac{\partial}{\partial R}
%\left[R^{3/2}\nue\Sig\frac{\partial v_{\rm R}}{\partial R}\right]
%-
%\frac{2}{3R^3}\frac{\partial(R^2\nue\Sig v_{\rm R})}{\partial R}
%\; ,
%\nonumber
%\end{eqnarray}

\begin{figure*}
\begin{center}
\rotatebox{270}{
\resizebox{65mm}{88.1mm}{
\mbox{
\includegraphics{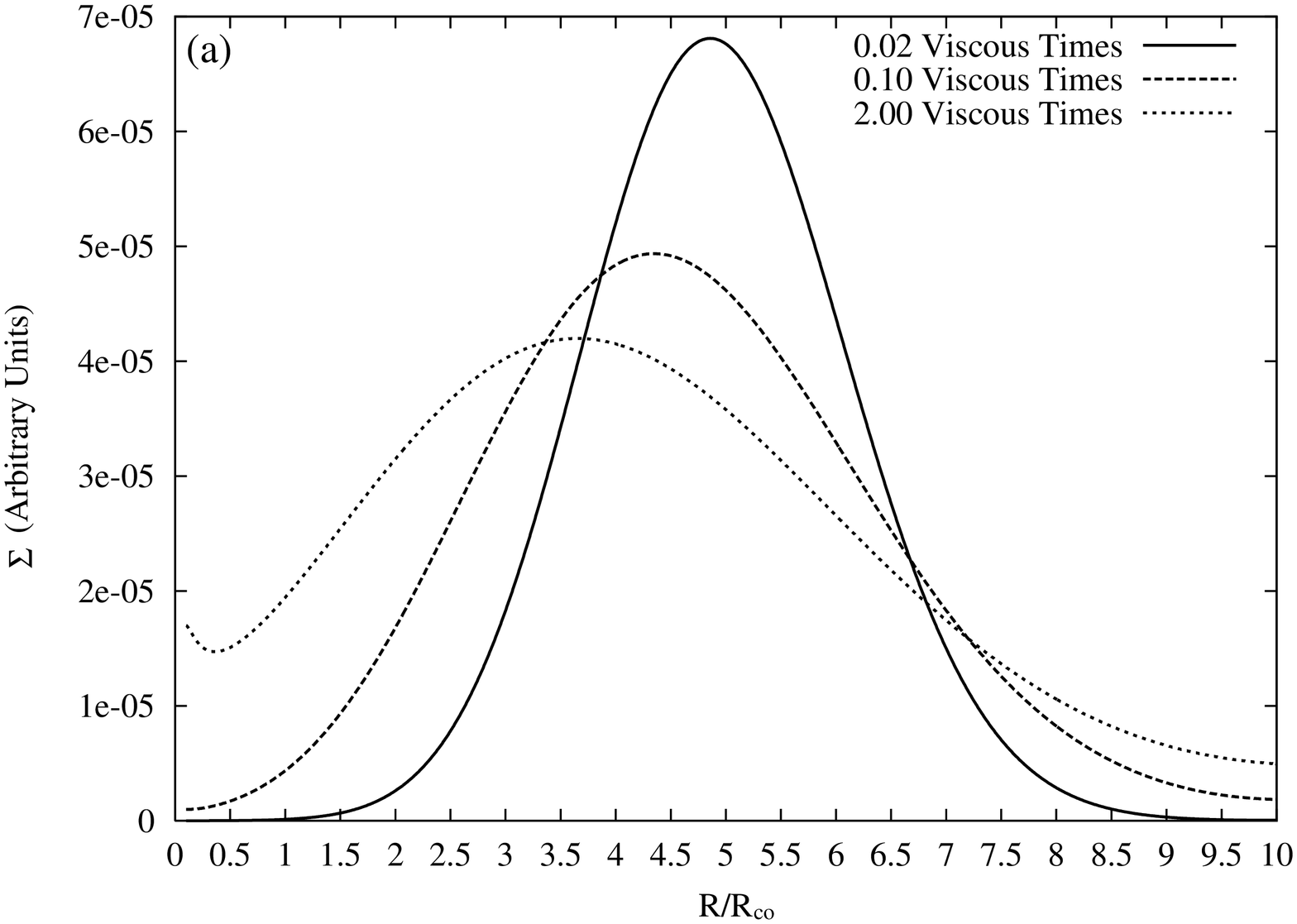}}}} 
\rotatebox{270}{
\resizebox{65mm}{88.1mm}{
\mbox{
\includegraphics{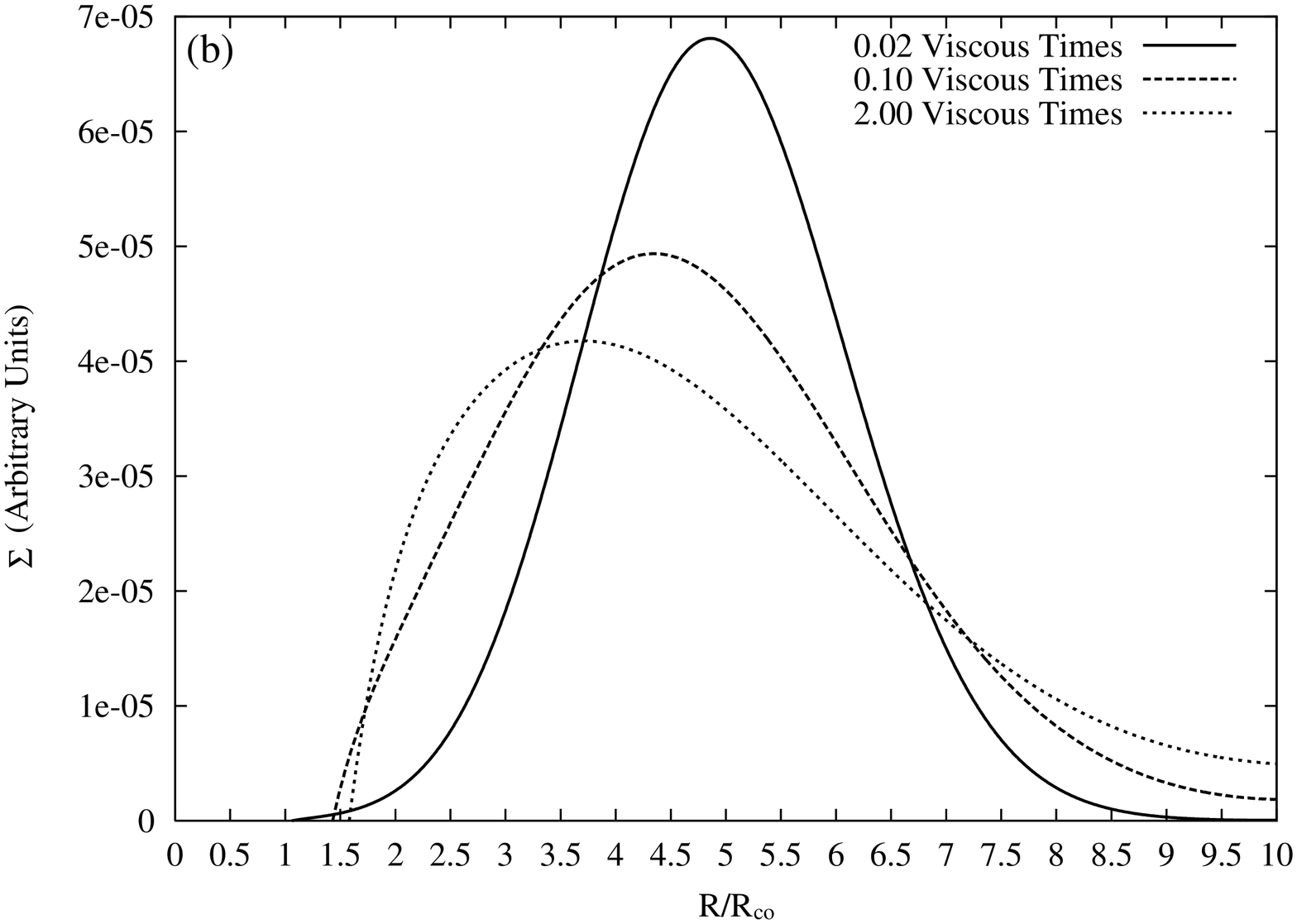}}}} 
\end{center}
\caption{Plot of surface density $\Sig$ (in code units) against radius for a spreading Gaussian disc with: (a) no advective term and; (b) a strong advective term which results from a magnetic field. The evolution of the depleted region or `hole' is clearly visible in the latter case.}
\label{ploto1}
\end{figure*}

\begin{figure*}
\begin{center}
\rotatebox{270}{
\resizebox{65mm}{88.1mm}{
\mbox{
\includegraphics{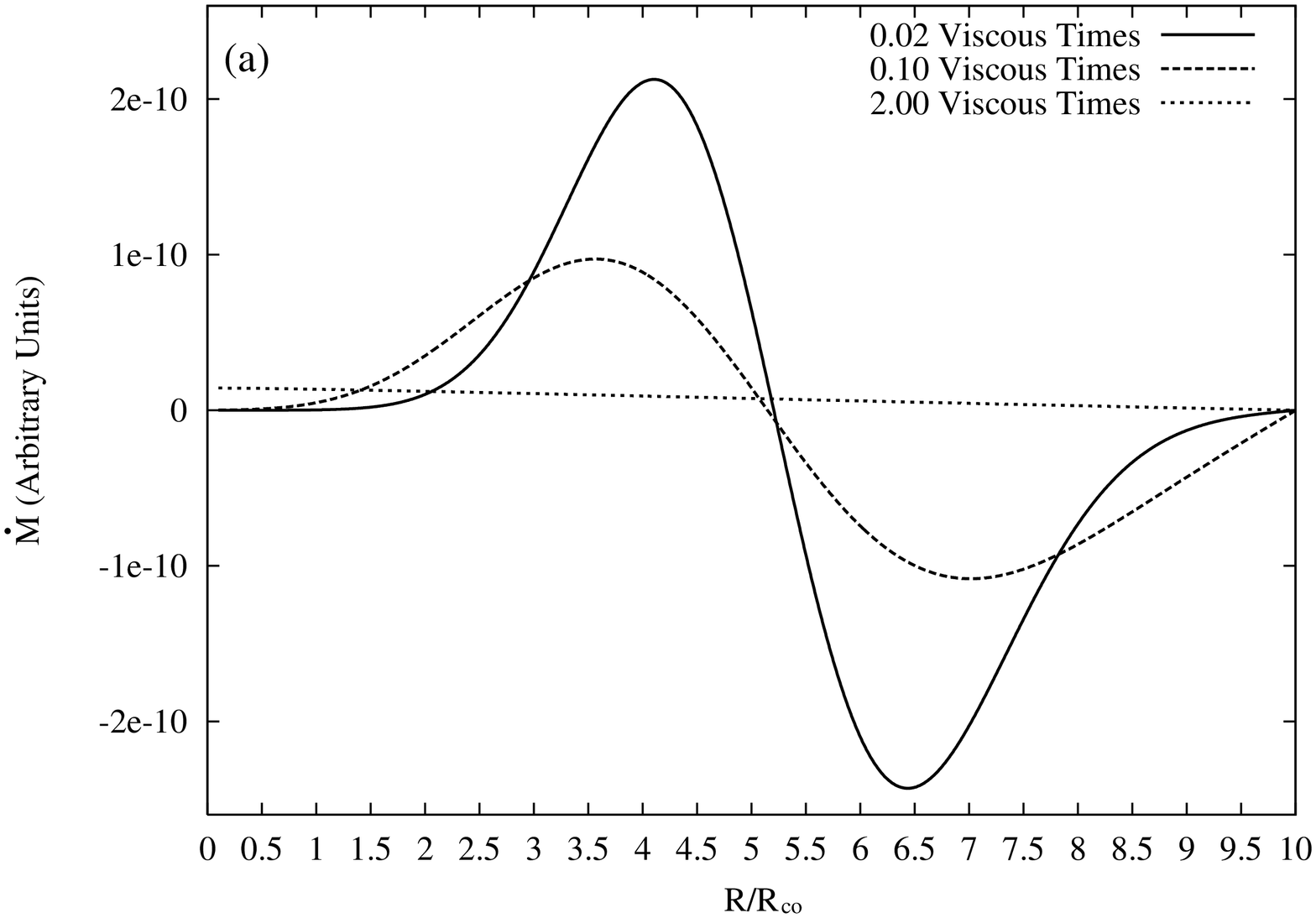}}}} 
\rotatebox{270}{
\resizebox{65mm}{88.1mm}{
\mbox{
\includegraphics{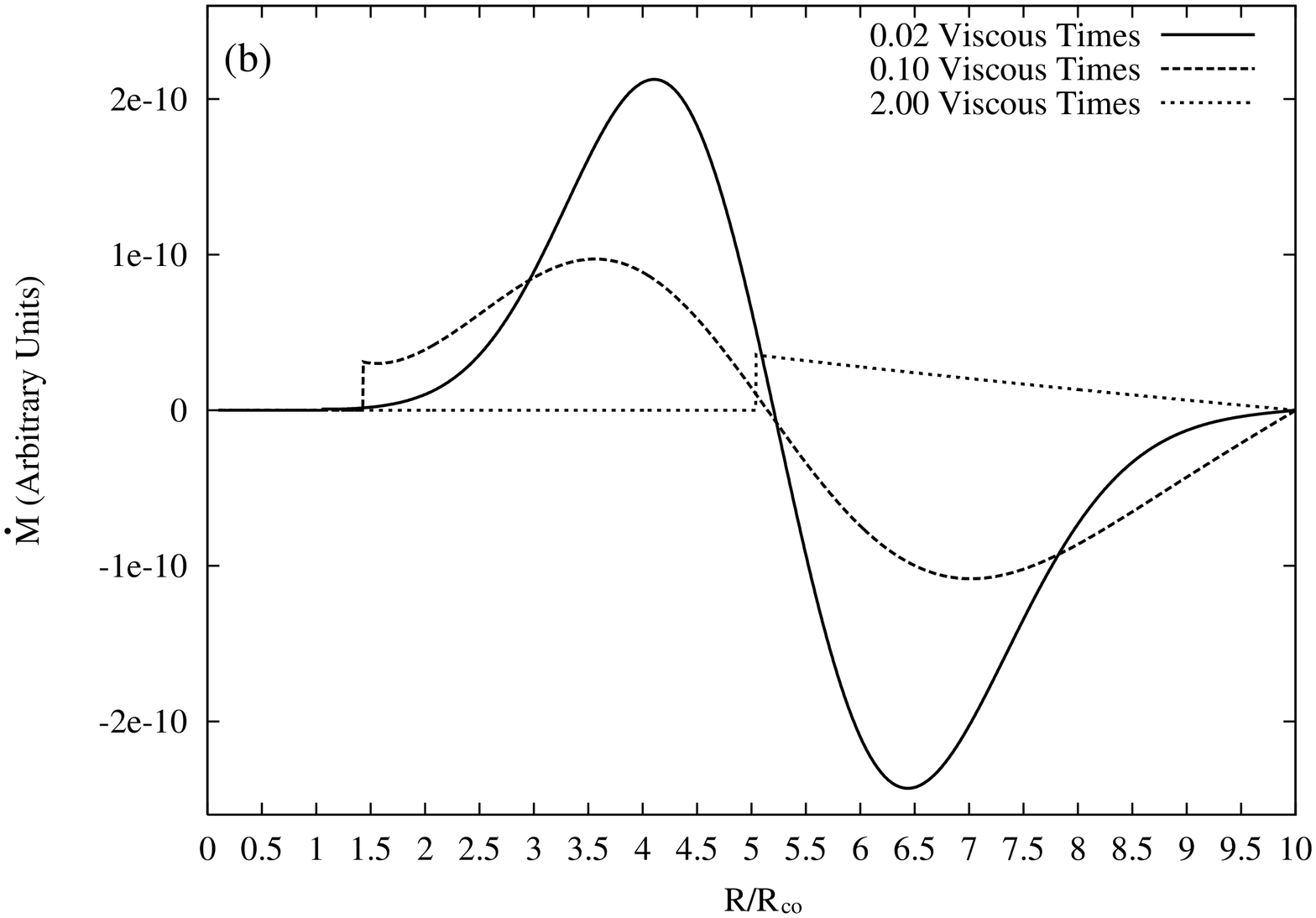}}}} 
\end{center}
\caption{Plot of mass transfer rate, $\mdot$ (in code units), against radius for a spreading Gaussian disc with: (a) no advective term and; (b) a strong advective term which results from a magnetic field. Where $\Sigma < 0$, $\mdot$ is not plotted. These are the same simulations, plotted at the same times as Figure \ref{ploto1}.} 
\label{ploto2}
\end{figure*}
 
\begin{figure*}
\begin{center}
\rotatebox{270}{
\resizebox{65mm}{88.1mm}{
\mbox{
\includegraphics{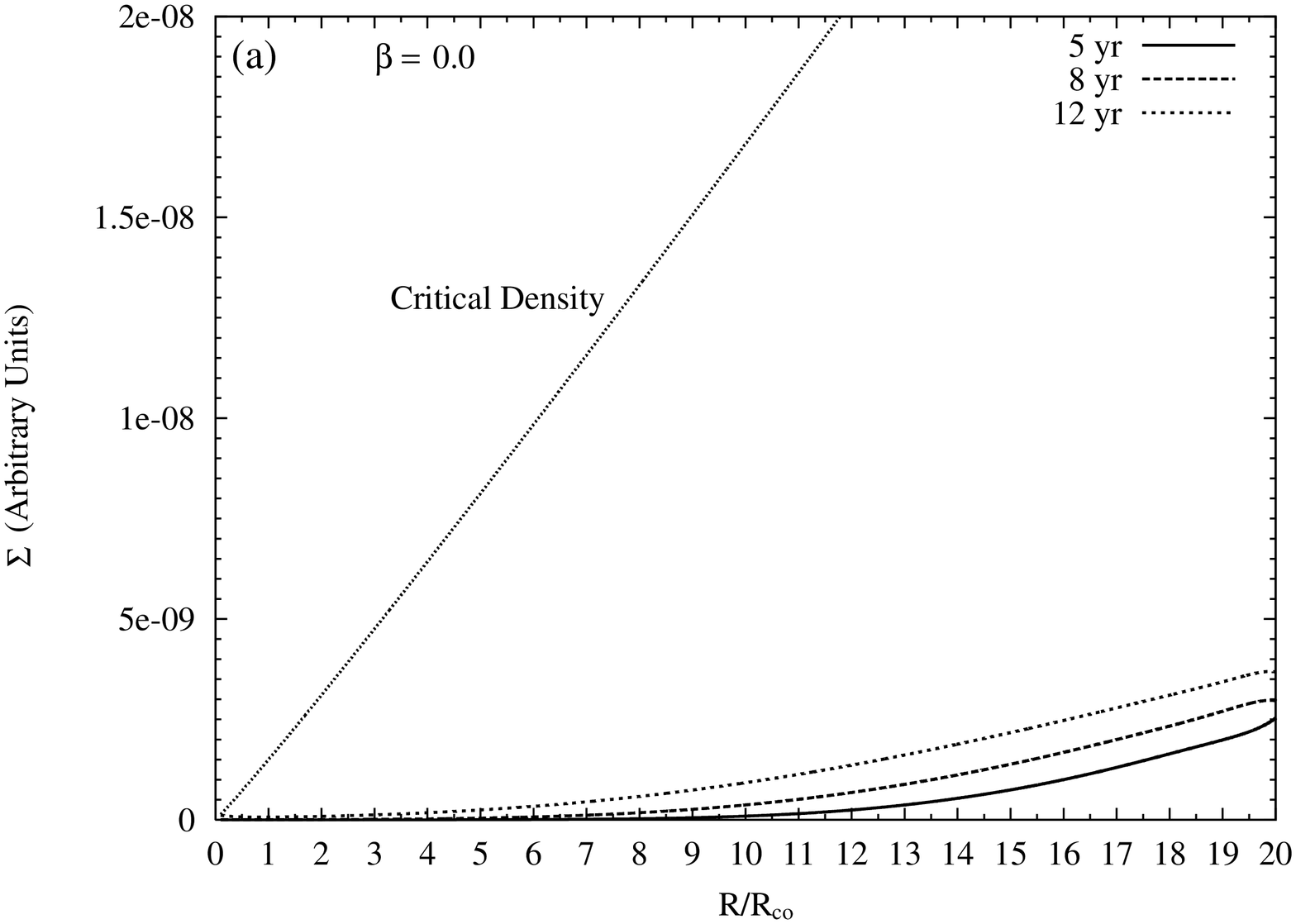}}}} 
\rotatebox{270}{
\resizebox{65mm}{88.1mm}{
\mbox{
\includegraphics{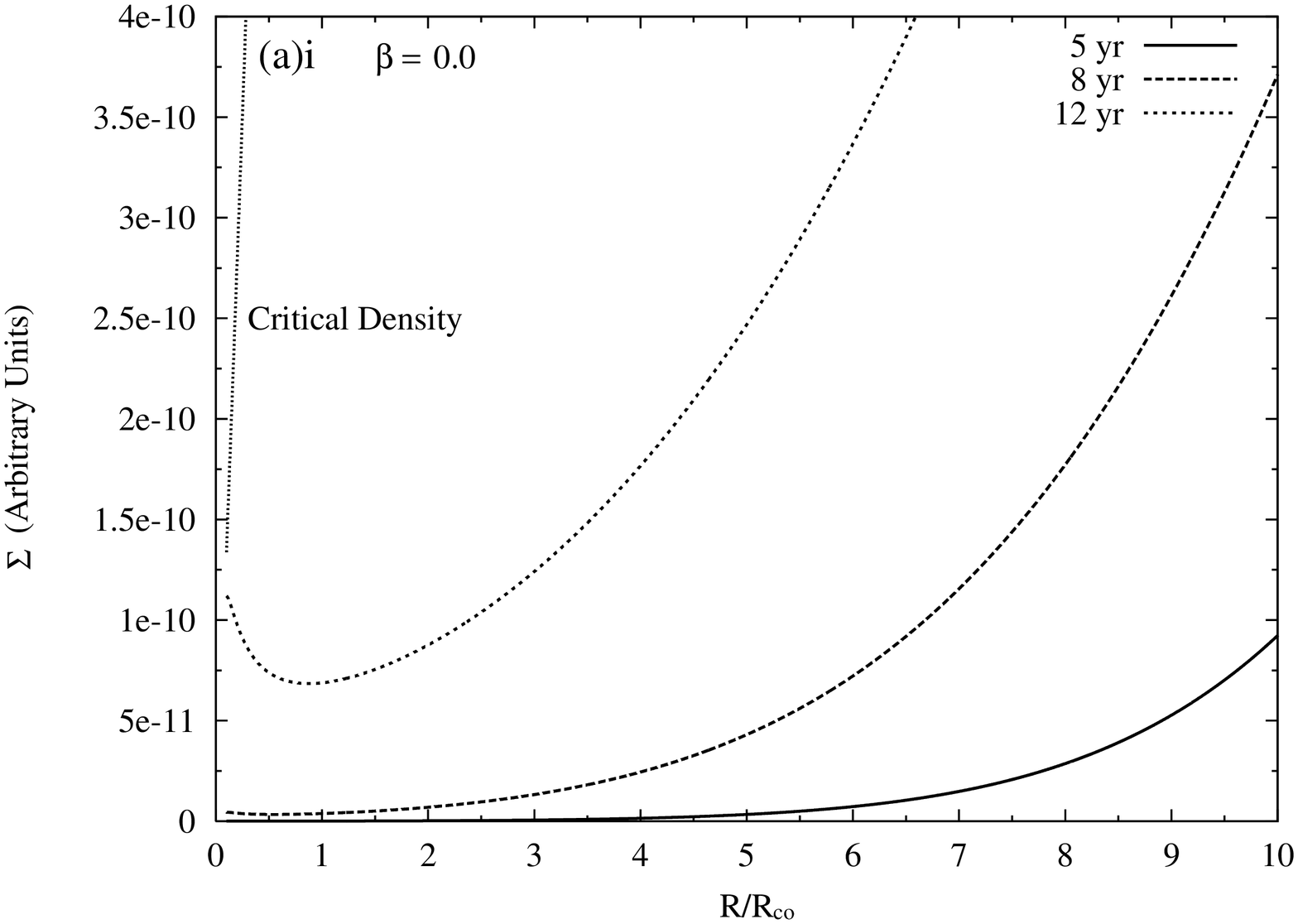}}}} 
\rotatebox{270}{
\resizebox{65mm}{88.1mm}{
\mbox{
\includegraphics{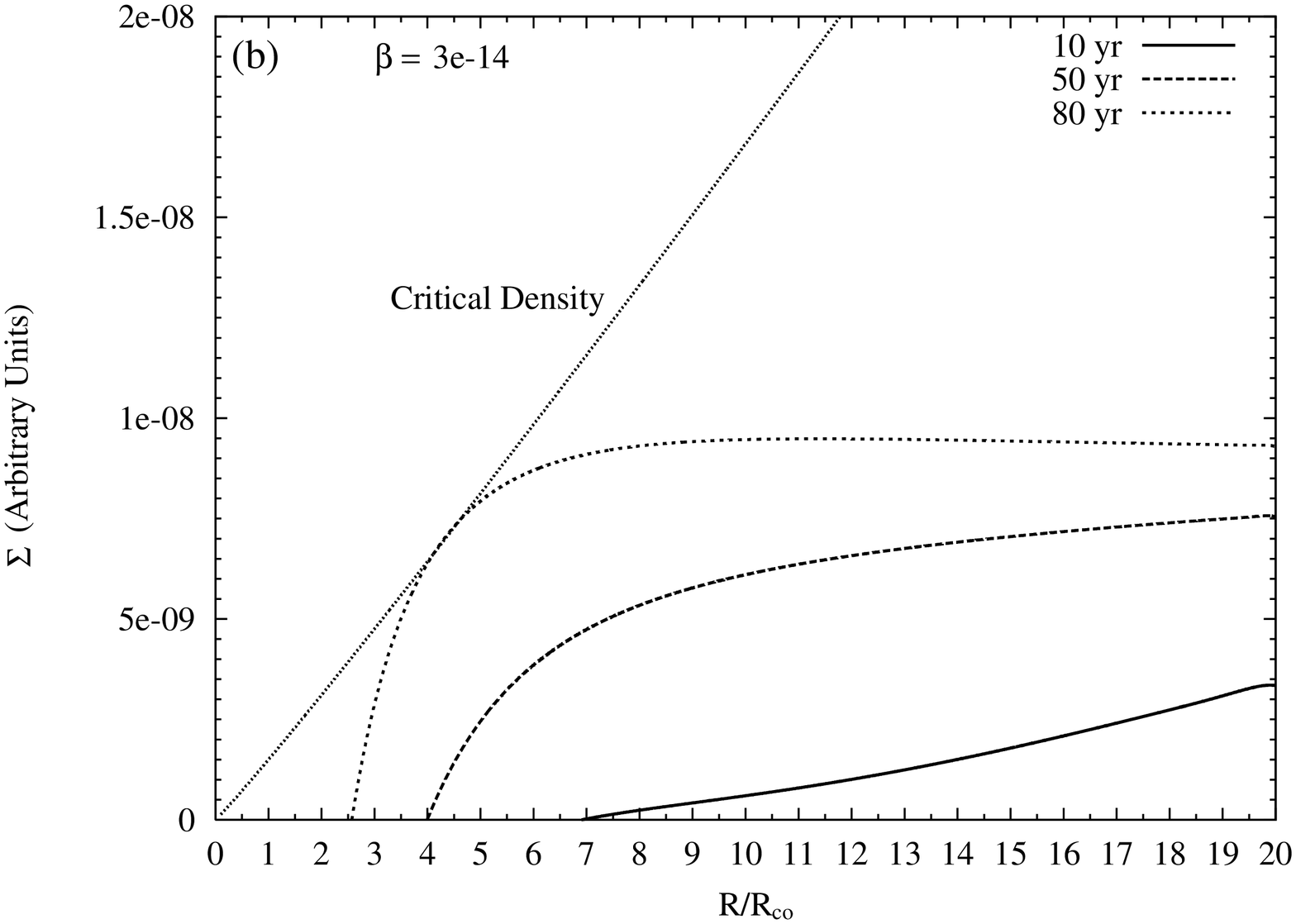}}}} 
\rotatebox{270}{
\resizebox{65mm}{88.1mm}{
\mbox{
\includegraphics{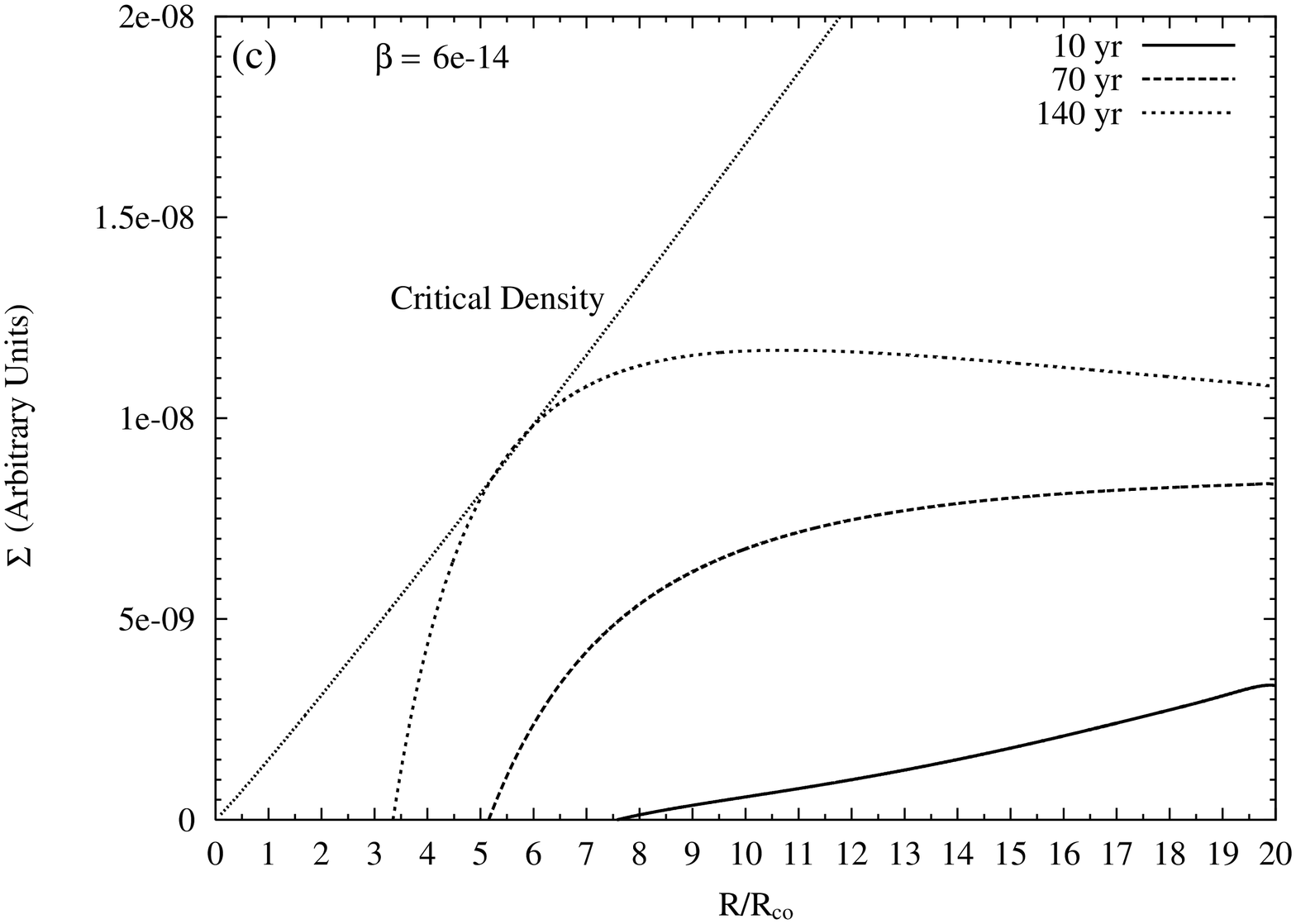}}}}
\end{center}
\caption{Plots of surface density $\Sig$ (in code units) against radius where advective terms of varying strengths, representing magnetic fields, have been applied. In all cases the initial conditions were an empty disc with a constant inflow of $\mdot \sim 10^{-7} \  {\rm{\msun yr{-1}}}$ from the right. Plot (a) shows the behaviour of a disc with a zero magnetic field while plots (b) and (c) represent a weak and a stronger field respectively. In both magnetic cases an inner disc truncation can clearly be seen, although this occurs at a greater radius in case (c). Plot (a)i is a `zoomed in' version of plot (a) showing where the density first exceeds $\sigcrith$, while plots (a), (b) and (c) are shown on identical axes for ease of comparison. In each case the latest plot is at the point where $\Sigma$ first reaches the critical density.}
\label{ploto3}
\end{figure*}

\begin{figure*}
\begin{center}
\rotatebox{270}{
%\resizebox{45mm}{58mm}{
\resizebox{65mm}{88.1mm}{
\mbox{
\includegraphics{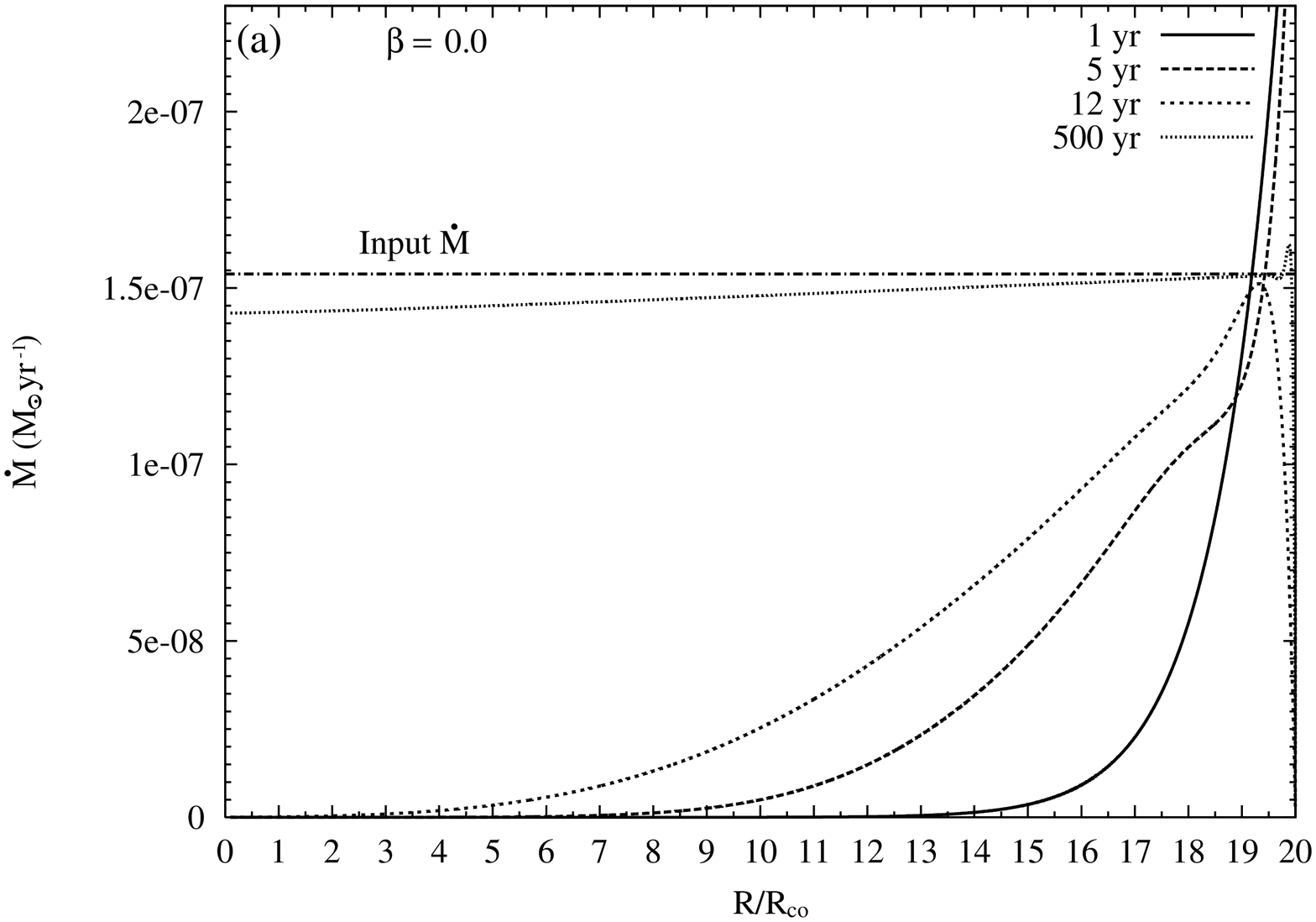}}}} 
\rotatebox{270}{
%\resizebox{45mm}{58mm}{
\resizebox{65mm}{88.1mm}{
\mbox{
\includegraphics{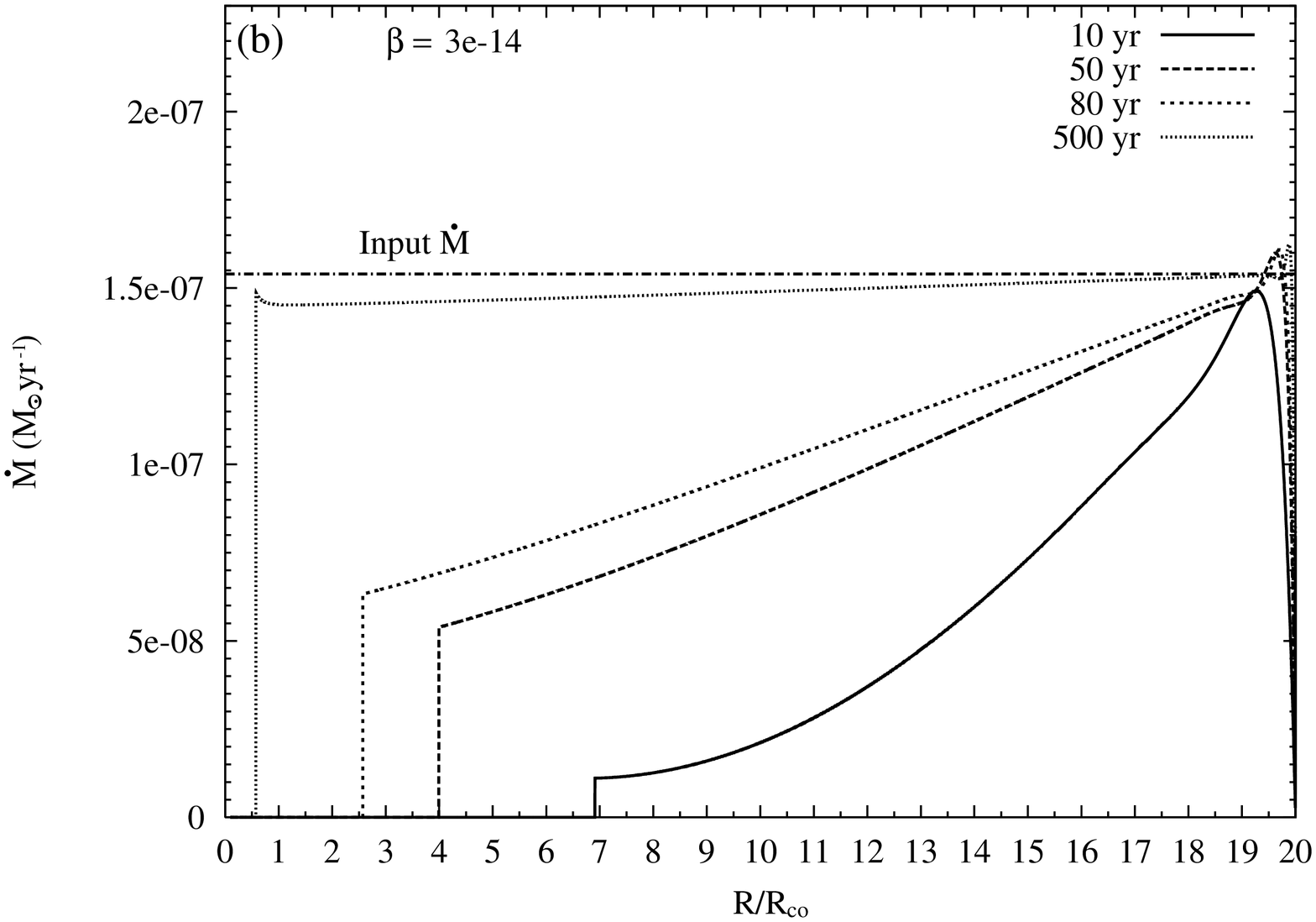}}}} 
\rotatebox{270}{
%\resizebox{45mm}{58mm}{
\resizebox{65mm}{88.1mm}{
\mbox{
\includegraphics{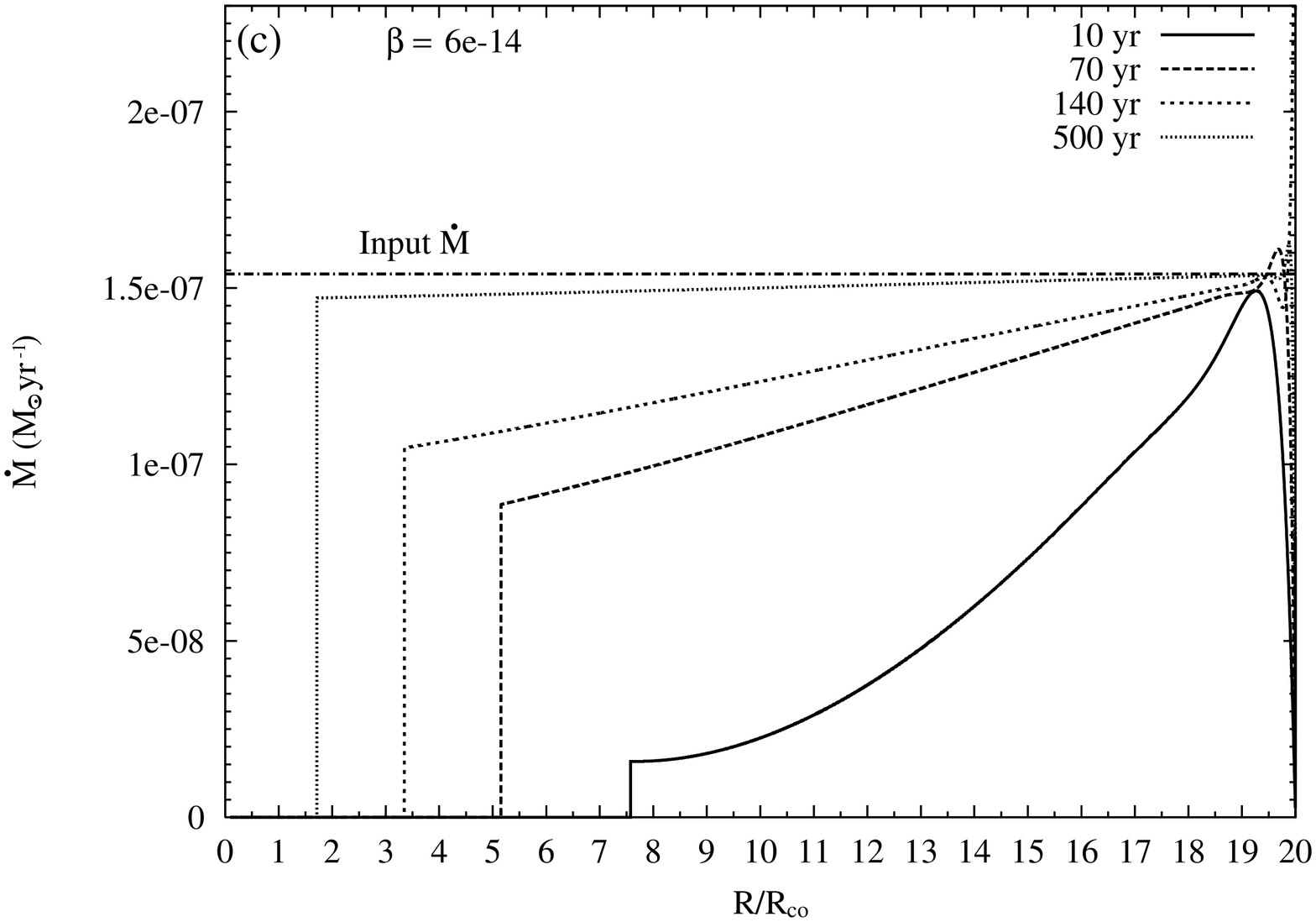}}}}
\end{center}
\caption{Plots of $\mdot$ (in code units) against radius for the same three advective terms strengths as were used in Figure \ref{ploto3}. The input $\mdot$ is plotted for comparison. Once again, where $\Sigma < 0$, $\mdot$ is not plotted. If no outburst occurred then the disc would eventually approach steady state. This can be seen in the plots at 500 yr, where the disc has been allowed to evolve without outburst.}
\label{ploto4}
\end{figure*}

\begin{figure*}
\begin{center}
\rotatebox{270}{
\resizebox{65mm}{88.1mm}{
\mbox{
\includegraphics{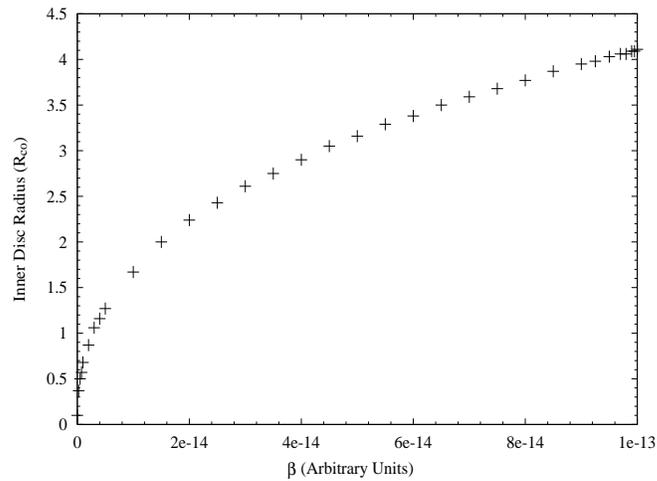}}}}
\end{center}
\caption{Plot of inner disc radius $\ri$ against magnetic parameter $\beta$ for a viscosity parameter of $\delta = 2 \times 10^{-8}$ and $\beta$ in code units. The curve can be fitted well by a power law of the form $\ri \propto \beta^{1/3}$.}
\label{ploto5}
\end{figure*} 

\begin{figure*}
\begin{center}
\rotatebox{270}{
\resizebox{65mm}{88.1mm}{
\mbox{
\includegraphics{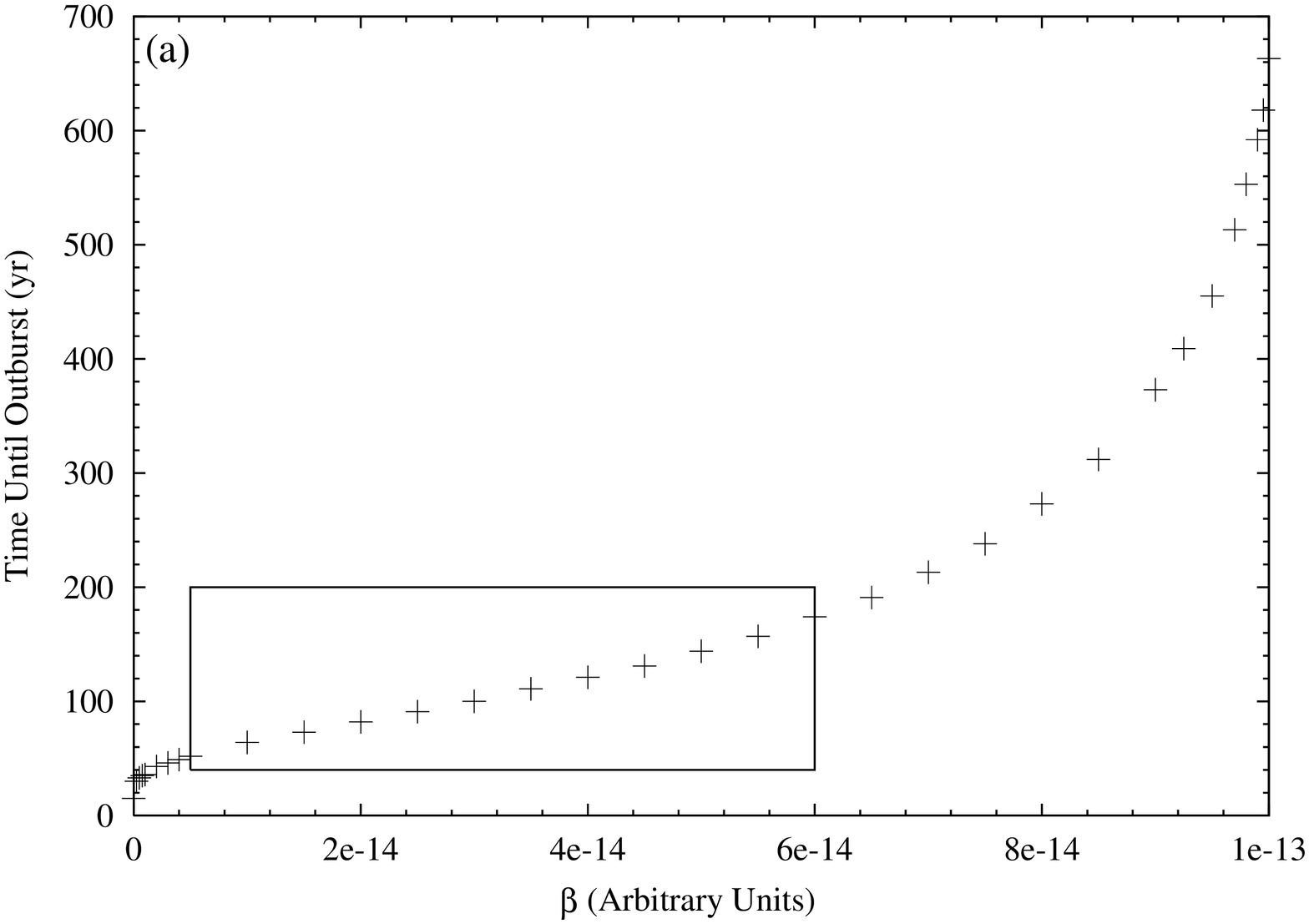}}}}
\rotatebox{270}{
\resizebox{65mm}{88.1mm}{
\mbox{
\includegraphics{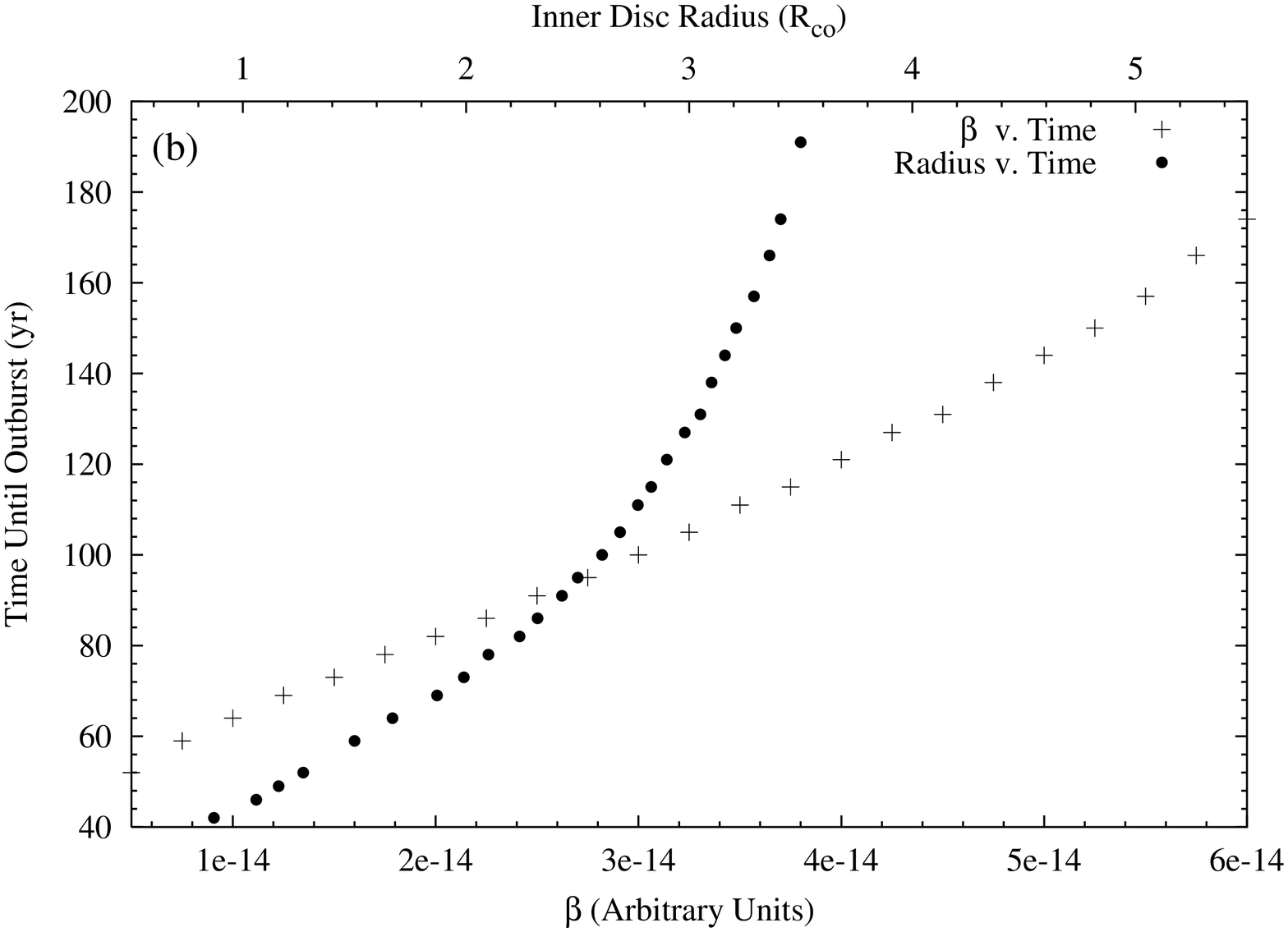}}}}
\end{center}
\caption{Plots of time before first outburst against magnetic parameter $\beta$. In plot (a) the time before the outburst increases very rapidly beyond a certain $\beta$. There, steady state is nearly reached before outburst. Plot (b) is a `zoomed in' view of plot (a), also indicating the corresponding inner disc radii.}
\label{ploto6}
\end{figure*}

\begin{figure*}
\begin{center}
\rotatebox{270}{
\resizebox{65mm}{88.1mm}{
\mbox{
\includegraphics{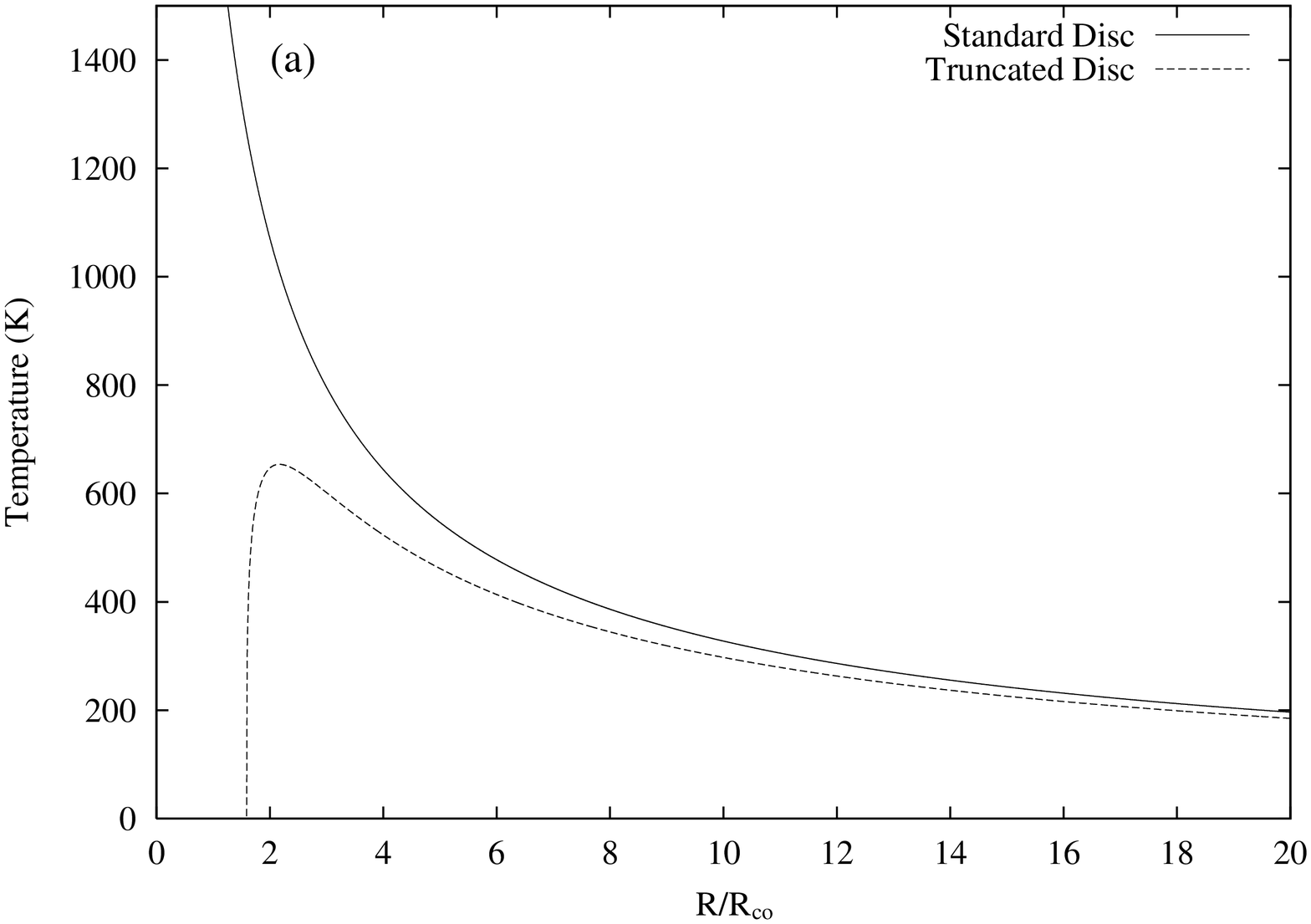}}}}
\rotatebox{270}{
\resizebox{65mm}{88.1mm}{
\mbox{
\includegraphics{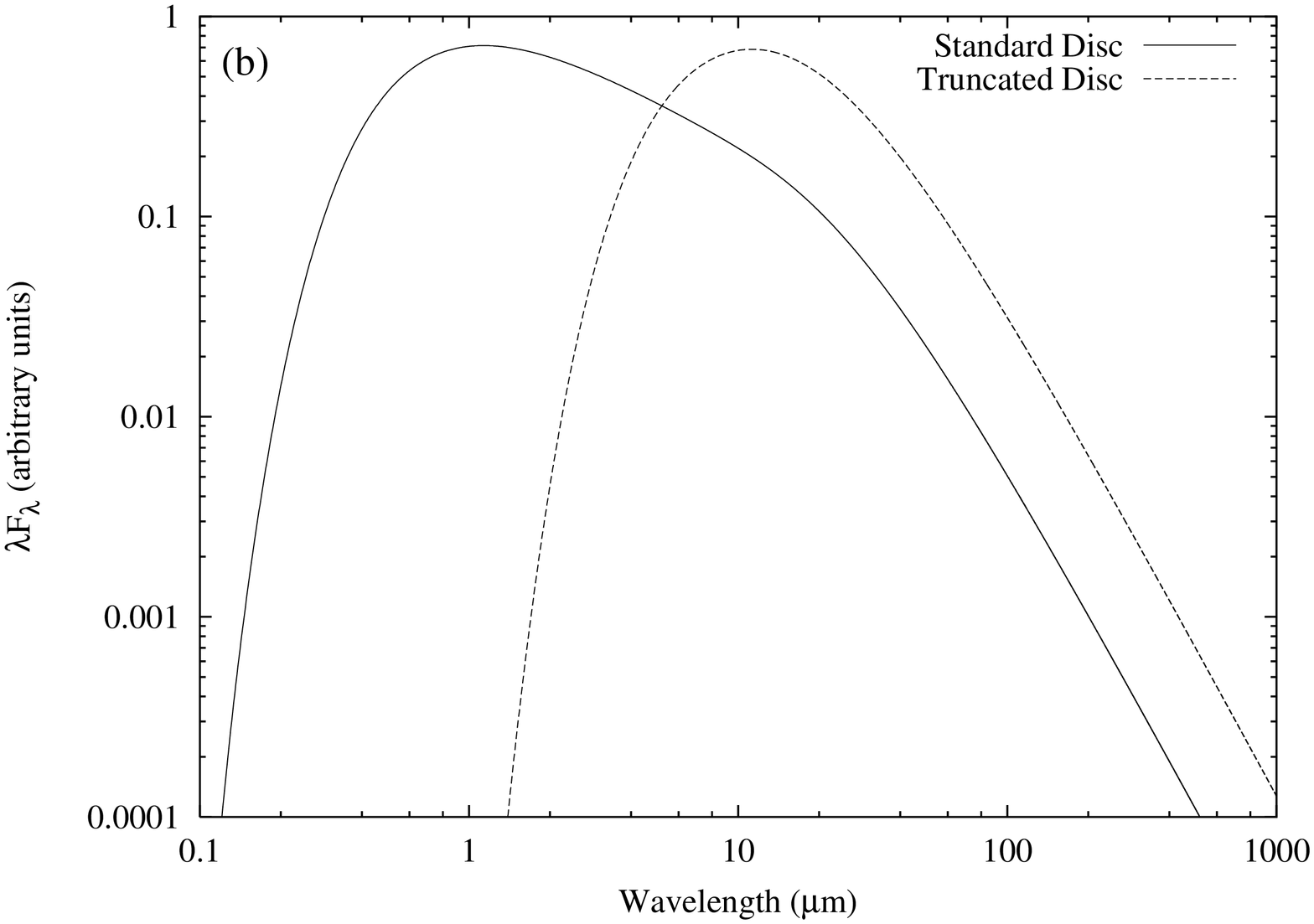}}}}
\end{center}
\caption{Plot (a) shows temperature in Kelvin as a function of radius in steady state disc models with and without magnetic truncation. The hottest part of the disc is missing in the truncated case. Plot (b) shows the spectral energy distribution in arbitrary units against wavelength in microns. In the truncated case the flux is reduced for wavelengths below $10 \mu{\rm m}$.}
\label{ploto7}
\end{figure*}

\subsection{Numerical Solutions}
\label{sec:res}
It will be shown by solving equation (\ref{sigmaevol}) numerically, that it is
feasible to create a depleted region of $\ri \sim 40 \ {\rm{\rsun}}$
in the centre of a T~Tau accretion
disc with a magnetic propeller and hence to increase $\trec$ to arbitrarily long times.
 The code is similar to that used by
\citet{arm96} to model a T Tau disc about a magnetic star, and their
results are similar to those presented below. The results are also comparable to those
of \citet{pri91}. In all cases the simulations tend towards a steady state which can be expressed
by the analytical solution derived above. The analytic solution has
been deliberately kept in as general
a form as possible, however for the numerical calculations $\epsilon =
3/4$ \citep{fra01} and $\gamma = 7/2$ have been adopted.
The computation is sensitive to resolution. 
Convergence with the analytic solution only occurs at high
resolution and when the inner boundary conditions match.

The solver code evaluates the diffusive and advective terms separately
and combines them using operator splitting. The diffusive term is
computed with a modified Crank-Nicholson scheme \citep{pre92} and the
resulting values of $\Sig$ are used as input for a modified two-step
Lax-Wendroff scheme \citep{pre92} which is used to solve the advective
term.

The first pair of simulations were performed starting from a Gaussian
density distribution, using the code described above. This initial
distribution represents a viscously spreading ring. One simulation was with, and
one without, an advection term, representing a magnetic field. The
parameters used in the code were as follows: the mass of the accreting
star was taken to be $\mstar = 1 \  \msun$. The co-rotation radius was
calculated on the basis of a star spinning with a typical T Tau period
of $3$ days which yields $\rco \sim 10$~$\rsun$.
 The boundary conditions applied at both the inner and
outer radii of the disc are outflow conditions and the initial density
distribution was centred at a radius of $5 \  \rco$. The disc was
simulated from $0.1 \  \rco \sim 1 \ {\rm{\rsun}}$ to $10 \
\rco$. The value $\beta \sim 10^{-13}$ is consistent with an
object of $R \sim 1 \  \rsun$ and $B \sim 1 \  {\rm{kG}}$. 

With the parameters described above, the effect of an advective term
on the surface density is illustrated in Figure~\ref{ploto1}. 
In the non-magnetic case,
where the diffusion term acts alone, the initial Gaussian mass
distribution quickly moves towards the accreting star, as
expected. This is very similar to the well known result of \citet{pri81}. In the
magnetic case, the evolution of the outer edge of the modelled region
is very similar to the non-magnetic case. This is not surprising
considering the $R^{-7/2}$ dependence of the advection term and is in
agreement with what was expected from the analytic solution. At the
inner edge however, it can clearly be seen that a depleted region of
radius $\sim 1.5$~$\rco$ is formed. 
It is interesting to observe how matter in the initial Gaussian
distribution is transported by the diffusion and advection
terms. Radial mass transfer rate $\mdot$ is plotted in Figure
\ref{ploto2}. The magnetic and non-magnetic cases appear quite similar
upon first inspection. Initially the movement of the
mass in the Gaussian is dominated by a diffusion effect. This is
represented by the two large lobes. However, in the non-magnetic case
the flux near the inner boundary increases in magnitude with
time. This represents an accretion process. In the magnetic case there
is no accretion, and the inward flux rapidly disappears with time and
the outward flux, driven by advection, is dominant at later times.     

Further simulations were performed using a constant inflow of $\mdot
\sim 10^{-7} \  {\rm{\msun yr^{-1}}}$ at
the outer boundary and an outflow condition at the inner boundary. In
this case an empty disc is assumed at the beginning of the
simulation. These conditions are qualitatively a more realistic
representation of the formation of the inner disc after an outburst,
and should more
correctly illustrate where mass build up would first trigger a new
outburst. Figure \ref{ploto3} shows the outcome of these
simulations with varying magnetic field strengths. The critical
density $\sigcrith$ is marked on these graphs to show the position in
the disc where surface density first exceeds $\sigcrith$ and outbursts
would begin. 

Figure \ref{ploto3}(a) represents a simulation of the disc with no
magnetic field. The inner radius of the simulation lies at the surface
of the star $\rstar \sim 1 \  \rsun$. 
Plot (a) is on the same scale as plots (b)
and (c) for ease of comparison, and plot (a)i is a zoomed in view of graph (a). 
It can be seen in plot (a) that a fairly flat surface density
profile is formed with an upturn in surface density developing towards
the surface of the star. The surface density rises with time and the
critical density is first reached at the surface of the star, as
expected.
 
Also in Figure \ref{ploto3} the effects of weaker and stronger
propellers can be seen in plots (b) and (c) respectively. In both cases
the early stages of disc evolution are not dissimilar to those in case
(a) for most of the disc. The surface
density increases with time and as it does so the effect of the
magnetic field becomes apparent. In case (b) the inflow of mass towards the
star is suppressed inside $\sim 2 \rco$ and the critical density is
first exceeded at $\sim 4 \rco$. This occurs much later than in plot
(a), as expected. 

In plot \ref{ploto3}(c) a stronger propeller is illustrated, where the
advection term is greater. The plot is qualitatively very similar to
plot (b) but the completely depleted region extends to $3 \rco$ and
the outburst would begin at $\sim 5 \rco \sim 50 \  \rsun$. Although a
full outburst cycle is not simulated and therefore it is not possible
to draw detailed conclusions about $\trec$, it can be seen that
outbursts would begin later with increasing $\beta$ in these
simulations. Values of $\ri$ of the expected order of magnitude can be
seen in Figure \ref{ploto3}. If the outbursting region has a width of
order a tenth of the radius at which the outburst begins, as assumed
in Section \ref{sec:met}, then the approximation that $\Sigma \sim
\sigcrit$ in this region is clearly a good one in all three cases. 

The mass transfer $\mdot$ in the disc for all these simulations is
plotted in Figure \ref{ploto4} at the same times as in Figure
\ref{ploto3}. In all cases the disc has yet to reach steady
state at the point where the outburst begins. If no outburst occurred, the
disc would eventually approach steady state. Because the spreading of 
the disc slows as it nears steady state, the longest recurrence times will be 
observed when the critical density is reached at, or shortly before, steady 
state. When $\beta$ is increased so that steady state is reached with the surface density maximum at larger radii then no outbursts will occur and the disc will remain quiescent indefinitely. Such a steady state without outbursts could exist for example, in our results, when $\beta > 1 \times 10^{-13}$. This is equivalent to a stellar magnetic field of $B > 1 \ {\rm kG}$ in a typical CTT disc with $\alphac \sim 0.01$.

It is possible to observe trends in both $\ri$ and $\trec$ as a
function of $\beta$. Several simulations were performed with different
values of the parameter $\beta$. As expected, the inner radius
increases with $\beta$. This increase can 
be fitted very well with a power law of index $\sim 1/3$. The results
are shown in Figure \ref{ploto5}.
 The values of $\ri$ obtained here are consistent with those
which are required in Section \ref{sec:met}. 

In Figure 6 we plot the relation between $\beta$ and time until first outburst, 
starting with an empty disc. In plot (a) the time before outburst increases rapidly beyond
a certain $\beta$. This is a result of the disc model approaching steady state. This means
that it is possible to obtain an arbitrarily long recurrence time provided the
star has a sufficiently strong magnetic moment. The value of $\beta$ at which this occurs
is a function of $\delta$ but the form remains the same. Plot (b) shows an enlarged section 
of plot (a). We additionally plot the corresponding inner disc radius. The relation between $\ri$ 
and the time before first outburst can be fitted by an expression of the form $t \propto \ri^{3}$ 
as would be expected from Equation (5). Further, when the time before outburst is long, the inner
radius of the disc is close to $\ri \sim 40 \ \rsun$ as predicted in Section 2. However these
results do not provide quantitative estimates of $\trec$ because the initial conditions are
so different from those just after an outburst. 

\subsection{Simulated spectra}
In order to compare the results of numerical models to observations it is useful to produce simulated spectra. This can be done in a simple way by dividing the disc in to a series of concentric annuli. The viscous dissipation in each annulus can then be approximated by \citep{fra01}
\begin{equation}
\label{dissip}
D\left(R\right) = \frac{9}{8} \nu \Sigma \frac{G \mstar}{R^{3}}
\; ,
\end{equation}
where $\nu$ is the Shakura-Sunyaev viscosity as given by Equation (\ref{tscale2}) and $\Sigma$ is taken from simulation results. Stellar irradiation may effect the disc temperature and hence the SED. Following the approach of \citet{ada86} the total luminosity of the disc can be broken down into components due to viscous dissipation and to disc irradiation. The viscous component can be approximated by \citep{fra01} 
\begin{equation}
\label{visccomp}
S_{\rm{v}} = \frac{G \mstar \mdot}{2 \rstar}
\; .
\end{equation}   
In the limit of a large disc, the component of disc luminosity due to irradiation can be approximated by \citep{ada86}
\begin{equation}
\label{irradcomp}
S_{\rm{i}} = \pi \rstar^{2} \sigma T_{\star}^{4}
\; ,
\end{equation}
where $T_{\star}$ denotes the temperature of the star. The two components can be compared using typical CTT parameters of $T_{\star} \sim 4000 \ {\rm{K}}$, $\mstar \sim 1 \ {\rm{\msun}}$, $\mdot \sim 10^{-7} \ {\rm{\mdot yr^{-1}}}$ and $\rstar \sim 1 \ {\rm{\rsun}}$. In this case, the component of disc luminosity due to irradiation is of the order of a few percent of that due to viscosity. It is therefore resonable to neglect irradiation in the simple model applied here. Considering viscous heating only and assuming that the disc is in thermal equilibrium, the temperature of each annulus can be calculated from the Stefan-Boltzmann law according to
\begin{equation}
\label{stefan}
D\left(R\right) = e \sigma T^{4}
\; ,
\end{equation}
where $e$ is emissivity (assumed to be unity), $\sigma$ is the Stefan-Boltzmann constant and $T$ represents the effective temperature. Temperature profiles, calculated using this simple model, can be seen in Figure \ref{ploto7}(a) for both the truncated and non-truncated disc models. A spectrum emitted from each annulus is then generated, assuming that the disc behaves as a black body. The spectral radiancy is then given by the Planck radiation law so that
\begin{equation}
\label{bb}
R_{\nu} = \frac{2 \pi h \nu^{3}}{c^{2}\left(e^{\frac{h \nu}{k T}}-1\right)}
\; ,
\end{equation}
where $h$ represents Planck's constant, $\nu$ is the frequency of the radiation, $k$ is Boltzmann's constant and $c$ is the speed of light. All the spectra of the annuli are summed to give the spectral energy distribution (SED) of the entire disc. A result is presented Figure \ref{ploto7}(b). In the case of the truncated disc the flux at wavelengths shorter than $1 \mu{\rm m}$ is greatly reduced. This is to be expected, as the hottest part of the disc has been significantly depleted. This may help to explain the infrared excess observed in some YSOs. The shape of the SED at longer wavelengths is not fully resolved as the contributions of the outer disc, beyond $200 \ \rsun$, have not been assessed. 

Observed spectral energy distributions of CTTs tend to include an infrared excess which is attributable, at least in part, to the presence of a circumstellar accretion disc \citep*{ber88,ada90}. \citet*{ito03} successfully reproduce observed SEDs for CTTs by summing the SED expected from the young stars themselves with disc SEDs. The position in wavelength of the inner edges, and peaks, of the disc SEDs used by \citet{ito03} lie between those of our truncated and non-truncated models which have been allowed to reach the steady state without reaching outburst. \citet{ult01} argue that a truncated accretion disc produces an infrared excess more closely in agreement with observations than an unmodified disc model. It should be noted that it is very difficult to separate the disc component from the dominant stellar component of the observed SEDs at wavelengths shorter than $10 \mu{\rm m}$. Detailed comparison with observation is problematic as the method used to generate disc SEDs from density profiles in this paper is only very approximate, and takes no account of material which is able to reach radii less than $\rmag$. Further comparison of observational results to the predictions of our model would therefore require a more detailed analysis than is possible under the restrictions of a one dimensional treatment.

%END OF SECTION

\section{Discussion}
\label{sec:con}

The FU Ori stars undergo what appear to be thermal-viscous
outbursts. It is very difficult however to reconcile observational
estimates of $\trec$ and  $\trs$ with the very
short estimates obtained from the standard thermal-viscous disc
instability  model with $\alphac \gtrsim 10^{-3}$  It is possible to do so by using a very low value of $\alphac$. We propose an alternative scheme in which outbursts occurring in an 
accretion disc truncated at a radius $\ri \sim 40 \  \rsun$ are 
consistent with the observational estimates of the outburst, recurrence,
and rise times. The central star, acting as a magnetic propeller, can 
produce the necessary truncation of the disc. The results of our
one dimensional model confirm the feasibility of this idea. 
A depleted region in the 
centre of the disc can be made large enough to produce $\trec$ similar 
to those estimated from observations. 
In particular, an arbitrarily long $\trec$ can be obtained
by selecting appropriate values for viscosity and 
magnetic field strength. 

The simulation of a full outburst in a refinement of the one
dimensional code described above, or using a three dimensional
technique such as smoothed particle hydrodynamics, would provide more accurate estimates of 
the outburst timescales, which could be compared to analytical results and to
observations. Three dimensional simulations would be particularly 
valuable as they would produce a self consistent picture of the inner
disc. It would also be desirable to treat viscosity in a more rigorous manner.
A more extensive study of stellar spin evolution would allow quantitative 
analysis of any possible feedback mechanisms.

The prolongation of recurrence times in accreting objects as a result of 
the magnetic propeller mechanism need not be confined to young stellar objects.
The same mechanism could potentially exist in some accreting binaries. In YSOs 
the full significance of the magnetic propeller may not yet be known. It is 
possible for example that a depleted inner disc may have an effect on the
planet formation process.  

\section*{ACKNOWLEDGEMENTS}
\label{sec:ack}
Research in theoretical astrophysics at the University of Leicester is
supported by a PPARC rolling grant. OMM gratefully acknowledges support
through a PPARC research studentship. Calculations were performed on
the Theoretical Astrophysics Group's PC cluster which is supported by
Advanced Micro Devices (AMD). We thank an anonymous referee for very 
helpful suggestions which helped to considerably clarify the paper.

% Further acknowledgments
% Star means no section number (or full width of page for figure}


\begin{thebibliography}{}
%If there are several references with the same first author, arrange
%in the following order: first single-author papers (by date); then
%two-author papers (alphabetically by co-author, then by date); then
%multi-author papers (by date). 
%\bibitem[Aconferee 2001] {ac01} Aconferee J., 2001, Proceedings of
%the International Symposium on Cheese, Leicester 11/2001
%\bibitem[Plonker 1966] {pl66} Plonker A., 1966, MNRAS, 999, 666
%\bibitem[Tom, Dick \& Harry 1066] {tdh66} Tom T., Dick D, Harry H.,
%1066, Introductory Plumbing, Saunders College Publishing
%\end{thebibliography}
%\When citing long use \citep{} (paretheses) or short use \citet{}
%(text), get full author list with * before the { for three author
%refs - that is not et al.

\bibitem[\protect\citeauthoryear{Adams \& Shu}{1986}]{ada86}Adams F.C., Shu F.H., 1986, ApJ, 308, 836
\bibitem[\protect\citeauthoryear{Adams, Lada \& Shu}{Adams \etal}{1987}]{ada87}Adams F.C., Lada C.J., Shu F.H., 1987, ApJ, 312, 788
\bibitem[\protect\citeauthoryear{Adams, Emerson \& Fuller}{Adams \etal}{1990}]{ada90}Adams F.C., Emerson J.P., Fuller G.A., 1990, ApJ, 357, 606
\bibitem[\protect\citeauthoryear{Armitage \& Clarke}{1996}]{arm96}Armitage P.J., Clarke C.J., 1996, MNRAS, 280, 458
%\bibitem[\protect\citeauthoryear{Basri, Marcy \& Valenti}{Basri \etal}{1992}]{bas92}Basri G., Marcy G.W., Valenti J.A., 1992, ApJ, 390, 622
%\bibitem[\protect\citeauthoryear{Bell \etal}{1995}]{bel95}Bell K.R., Lin D.N.C., Hartmann L.W., Kenyon S.J., 1995, ApJ, 444, 376
\bibitem[\protect\citeauthoryear{Beckwith \etal}{1984}]{bec84}Beckwith S., Skrutskie M.F., Zuckerman B., Dyck H.M., 1984, ApJ, 287, 793
\bibitem[\protect\citeauthoryear{Bell \& Lin}{1994}]{bel94}Bell K.R., Lin D.N.C., 1994, ApJ, 427, 987
\bibitem[\protect\citeauthoryear{Bell \etal}{1995}]{bel95}Bell K.R., Lin D.N.C., Hartmann L.W., Kenyon S.J., 1995, ApJ, 444, 376
\bibitem[\protect\citeauthoryear{Bertout, Basri \& Bouvier}{Bertout \etal}{1988}]{ber88}Bertout C., Basri G., Bouvier J., 1988, ApJ, 330, 350
\bibitem[\protect\citeauthoryear{Bouvier \etal}{1995}]{bou95}Bouvier J., Covino E., Kovo O., Martin E.L., Matthews J.M., Terranegra L., Beck S.C., 1995, A\&A 299, 89
\bibitem[\protect\citeauthoryear{Buat-M{\'e}nard, Hameury \& Lasota}{Buat-M{\' e}nard \etal}{2001}]{bua01}Buat-M{\' e}nard V., Hameury J.M., Lasota J.P., 2001, A\&A, 366, 612
\bibitem[\protect\citeauthoryear{Cannizzo, Shafter \& Wheeler}{Cannizzo \etal}{1988}]{can88}Cannizzo J.K, Shafter R.A., Wheeler J.C., 1988, ApJ, 333, 227
\bibitem[\protect\citeauthoryear{Clarke \& Syer}{1996}]{cla96}Clarke C.J., Syer D., 1996, MNRAS, 278, L23
\bibitem[\protect\citeauthoryear{Clarke \etal}{1995}]{cla95}Clarke C.J., Armitage P.J., Smith K.W., Pringle J.E., 1995, MNRAS, 273, 639 
\bibitem[\protect\citeauthoryear{Dendy}{1990}]{den90}Dendy R.O., 1990, Plasma Dynamics, Oxford University Press, Oxford
\bibitem[\protect\citeauthoryear{Frank, King \& Raine}{Frank \etal}{2002}]{fra01}Frank J., King A.R., Raine D.J., 2002, Accretion Power in Astrophysics, 3rd edn. Cambridge Univ. Press, Cambridge 
\bibitem[\protect\citeauthoryear{Gammie}{1996}]{gam96}Gammie C.F., 1996, ApJ, 457, 355
%\bibitem[\protect\citeauthoryear{Guenther}{1997}]{gue97}Guenther E.W., 1997, in Reipurth B., Bertout C., eds, Proc IAU Symp. 182, Herbig-Haro Flows and the Birth of Low-Mass Stars. Kluwer, Dordrecht, p. 433
\bibitem[\protect\citeauthoryear{Gullbring \etal}{1998}]{gul98}Gullbring E., Hartmann L., Brice{\~n}o C., Calvet N., 1998, ApJ, 492, 323
\bibitem[\protect\citeauthoryear{Hartmann}{1998}]{har98}Hartmann L., 1998, Accretion Processes in Star Formation, Cambridge Univ. Press, Cambridge
\bibitem[\protect\citeauthoryear{Hartmann \& Kenyon}{1985}]{har85}Hartmann L., Kenyon S.J., 1985, ApJ, 299, 462
%\bibitem[\protect\citeauthoryear{Hartmann \& Kenyon}{1988}]{har88}Hartmann L., Kenyon S.J., 1988, ApJ, 312, 243
\bibitem[\protect\citeauthoryear{Hartmann \& Kenyon}{1996}]{har96}Hartmann L., Kenyon S.J., 1996, ARA\&A, 34, 207 
\bibitem[\protect\citeauthoryear{Hartmann \etal}{1998}]{har97}Hartmann L., Calvet N., Gullbring E., D'Alessio P., 1998, ApJ, 495, 385
\bibitem[\protect\citeauthoryear{Itoh \etal}{2003}]{ito03}Itoh I., Sugitani K., Fukuda N., Nakanishi K., Ogura K., Tamura M., Marui K., Fujita K., Oasa Y., Fukagawa M., 2003, ApJ, 586, L141
\bibitem[\protect\citeauthoryear{Johns-Krull \etal}{1999}]{joh99}Johns-Krull C.M., Valenti J.A., Hatzes A.P., Kanaan A., 1999, ApJ, 510, L39
\bibitem[\protect\citeauthoryear{Johnstone \& Penston}{1987}]{joh87}Johnstone R.M., Penston M.V., 1987, MNRAS, 227, 797
\bibitem[\protect\citeauthoryear{Kenyon \& Hartmann}{1987}]{ken87}Kenyon S.J., Hartmann L., 1987, ApJ, 323, 733  
\bibitem[\protect\citeauthoryear{Kenyon, Hartmann \& Hewett}{Kenyon \etal}{1988}]{ken88}Kenyon S.J., Hartmann L., Hewett R., 1988, ApJ, 325, 231
%\bibitem[\protect\citeauthoryear{Kenyon, Calvet \& Hartmann}{Kenyon \etal}{1993}]{ken93}Kenyon S.J., Calvet N., Hartmann L., 1993, ApJ, 414, 676
\bibitem[\protect\citeauthoryear{Kenyon, Insu \& Hartmann}{Kenyon \etal}{1996}]{ken95}Kenyon S.J., Insu Y., Hartmann L., 1996, ApJ, 462, 439 
\bibitem[\protect\citeauthoryear{Lin \& Papaloizou}{1986}]{lin86}Lin D.N.C., Papaloizou J., 1986, MNRAS, 309, 846
\bibitem[\protect\citeauthoryear{Livio \& Pringle}{1992}]{liv92}Livio M., Pringle J.E., 1992, MNRAS, 259, 23P
\bibitem[\protect\citeauthoryear{Lovelace, Romanova \& Bisnovatyi-Kogan}{Lovelace \etal}{1999}]{lov99}Lovelace R.V.E., Romanova M.M., Bisnovatyi-Kogan G.S., 1999, ApJ, 514, 368
%\bibitem[\protect\citeauthoryear{Machado, Lago \& Lima}{Machado \etal}{1999}]{mac99}Machado L.J.R., Lago M.T.V.T., Lima J.J.G. 1999, Ap\&SS, 261, 69
\bibitem[\protect\citeauthoryear{Murray \etal}{1999}]{mur99}Murray J.R., Armitage P.J., Ferrario L., Wickramasinghe D.T., 1999, MNRAS, 302, 189
\bibitem[\protect\citeauthoryear{Pearson, Wynn \& King}{Pearson \etal}{1997}]{pea97}Pearson K.J., Wynn G.A., King A.R., MNRAS, 1997, 288, 421 
\bibitem[\protect\citeauthoryear{Popham}{1996}]{pop96}Popham R., 1996, ApJ, 467 749
\bibitem[\protect\citeauthoryear{Press \etal}{1992}]{pre92}Press W.H., Teukolsky S.A., Vetterling W.T., Flannery B.P., 1992, Numerical Recipes in Fortran, Cambridge Univ. Press, Cambridge
\bibitem[\protect\citeauthoryear{Pringle}{1981}]{pri81}Pringle J.E., 1981, ARA\&A, 19, 137
\bibitem[\protect\citeauthoryear{Pringle}{1991}]{pri91}Pringle J.E., 1991, MNRAS, 248, 754
%\bibitem[\protect\citeauthoryear{Rydgen \& Zak}{1987}]{ryd87}Rydgen A.E., Zak S.K., PASP, 99, 141
\bibitem[\protect\citeauthoryear{Safier}{1999}]{saf99}Safier P.N., 1999, ApJ, 510, L127
%\bibitem[\protect\citeauthoryear{Shevchenko \& Herbst}{1998}]{she98}Shevchenko V.S., Herbst W., 1998, AJ, 116, 1419
%\bibitem[\protect\citeauthoryear{Truss, Murray \& Wynn}{2001}]{tru01}Truss M.R., Murray J.R., Wynn G.A., 2001, MNRAS, 324, L1
\bibitem[\protect\citeauthoryear{Ultchin, Wynn \& Regev}{Ultchin \etal}{2001}]{ult01}Ultchin Y., Regev O., Wynn G.A., 2002, MNRAS, 331, 578  
\bibitem[\protect\citeauthoryear{Warner}{1995}]{war95}Warner B., 1995, Cataclysmic Variable Stars, Cambridge Univ. Press, Cambridge
%\bibitem[\protect\citeauthoryear{Warner, Livio \& Tout}{1996}]{war96}Warner B., Livio M., Tout C.A., 1996, MNRAS, 282, 735
%\bibitem[\protect\citeauthoryear{Wynn \& King}{1995}]{wyn95}Wynn G.A., King A.R., 1995, MNRAS, 275, 9W
\bibitem[\protect\citeauthoryear{Wynn, King \& Horne}{Wynn \etal}{1997}]{wyn97}Wynn G.A., King A.R., Horne K., 1997, MNRAS, 286, 436
%\bibitem[\protect\citeauthoryear{Wynn \etal}{2002}]{wyn02}Wynn G.A., Leach R., Matthews O.M., Speith R., King A.R., West R., 2002, MNRAS, submitted
%\bibitem[\protect\citeauthoryear{Yi}{1995}]{yi95}Yi I., 1995, ApJ, 442, 768
\bibitem[\protect\citeauthoryear{Yi \& Kenyon}{1997}]{yi97} Yi I., Kenyon S.J., 1997, ApJ, 477, 379
\end{thebibliography}
\end{document}